\documentclass[aps,prd,twocolumn,showpacs,10pt,superscriptaddress,preprintnumbers,nofootinbib]{revtex4-1}
\usepackage[utf8]{inputenc}
\usepackage{amsmath,amssymb,bm,slashed,braket}
\usepackage{graphicx}
\usepackage{float}
\usepackage[dvipsnames,table,svgnames]{xcolor}
\usepackage[normalem]{ulem}
\usepackage{colortbl}
\usepackage{multirow,boldline, booktabs}
\usepackage{makecell}
\usepackage{hhline}
\usepackage{dsfont}
\usepackage{verbatim}
\usepackage{MnSymbol}
\usepackage[colorlinks=true,
            linkcolor=blue,
            urlcolor=blue,
            citecolor=teal,          
            bookmarks=true,
            bookmarksnumbered=true,
            breaklinks=true,
            pdfpagemode=Fullscreen,
            pdfstartview=FitBH]{hyperref}
\allowdisplaybreaks[4]
\usepackage[capitalise]{cleveref}
\usepackage{orcidlink}
\usepackage{pifont}
\usepackage{cancel}
\usepackage{tikz} 
\usepackage{tkz-euclide}
\usetikzlibrary{backgrounds} 
\usetikzlibrary{decorations.pathmorphing}
\usetikzlibrary{arrows.meta}
\usetikzlibrary{shapes.misc}
\tikzset{
mystyle/.style={line width=1, baseline, scale=0.6, every node/.style={scale=1}},
photon/.style={decorate, decoration={snake, segment length=1.5 mm, amplitude=0.5mm}, draw=black, thick},
v/.style={decorate, draw, decoration={snake, segment length=2.mm, amplitude=0.5mm}},
f/.style={draw, decoration={markings,mark=at position #1 with {\arrow[]{Latex[length=1.5mm,width=1.5mm]}}},
    postaction={decorate},node contents=#1},
f/.default=.6,
fb/.style={draw,decoration={markings,mark=at position #1 with {\arrowreversed[]{Latex[length=1.5mm,width=1.5mm]}}},
    postaction={decorate},node contents=#1},
fb/.default=.6,
s/.style={dashed,draw, postaction={decorate},
        decoration={markings,mark=at position .7 with {\arrow[very thick]{latex}}}},
sb/.style={dashed,draw, postaction={decorate},
        decoration={markings,mark=at position .55 with {\arrowreversed[draw=black,very thick]{latex}}}},
snar/.style={dashed,draw,line width =1.25pt},
gluon/.style={decorate,
 decoration={coil,amplitude=2pt, segment length=3.5pt,  pre length=.1cm, post length=.1cm}},
}

\newcommand{\calO}{\mathcal{O}}
\newcommand{\N}{ {\tt N} }
\newcommand{\C}{ {\tt C} }
\newcommand{\tL}{ {\tt L} }
\newcommand{\tR}{ {\tt R} }

\begin{document}

\title{Renormalization-group-improved constraints on dimension-7 baryon-number-violating operators}
\author{Yi Liao\,\orcidlink{0000-0002-1009-5483}}
\email{liaoy@m.scnu.edu.cn}
\author{Xiao-Dong Ma\,\orcidlink{0000-0001-7207-7793}}
\email{maxid@scnu.edu.cn}
\author{Xiang Zhao\,\orcidlink{0009-0008-6024-7722}}
\email{zhaox@m.scnu.edu.cn}
\affiliation{State Key Laboratory of Nuclear Physics and
Technology, Institute of Quantum Matter, South China Normal
University, Guangzhou 510006, China}
\affiliation{Guangdong Basic Research Center of Excellence for
Structure and Fundamental Interactions of Matter, Guangdong
Provincial Key Laboratory of Nuclear Science, Guangzhou
510006, China}

\begin{abstract}
We study constraints on dimension-7 SMEFT baryon-number-violating  operators from nucleon decays by incorporating full renormalization group (RG) running effects. At high new physics scales, we demonstrate that RG running effects help set stringent bounds on all 297 Wilson coefficients compared to the tree-level analysis in which only coefficients involving the first and second fermion generations could be constrained.
Our findings highlight that the RG running effects through Yukawa mixings are particularly important for indirectly probing operators involving the second and third generation fermions.  
\end{abstract}

\maketitle 
%%%%%%%%%%%%%%%%%%%%%%%%
\section{ Introduction}
%%%%%%%%%%%%%%%%%%%%%%%%

Baryon number violation (BNV) is predicted in numerous scenarios beyond the standard model, ranging from high-scale grand unified theories~\cite{Pati:1974yy,Pati:1973uk,Fritzsch:1974nn,Georgi:1974sy,Nath:2006ut} to relatively lower-scale models involving leptoquarks~\cite{Buchmuller:1986zs,Dorsner:2016wpm,Assad:2017iib}. 
As a key ingredient in baryogenesis, scenarios with explicit BNV interactions could provide a potential explanation for the observed matter-antimatter asymmetry of the Universe.
The most feasible experimental approach to test these scenarios is the search for nucleon decays, leveraging the vast number of nucleons available in terrestrial detectors. 
Over the past few decades, experiments such as Frejus~\cite{Frejus:1991ben}, 
IMB~\cite{Irvine-Michigan-Brookhaven:1983iap,McGrew:1999nd}, 
Super-Kamiokande (Super-K)~\cite{Super-Kamiokande:2014pqx,Super-Kamiokande:2020tor}, and SNO+~\cite{SNO:2018ydj,SNO:2022trz} have conducted extensive searches for nucleon decays, establishing stringent limits on their occurrence~\cite{ParticleDataGroup:2024cfk}. 

Being agnostic about the underlying physics, the standard model effective field theory (SMEFT) offers a systematic framework to parametrize heavy new physics, particularly well-suited for BNV nucleon decay processes given their lower energy scale~\cite{Weinberg:1979sa,Wilczek:1979hc,Weinberg:1980bf}. 
In SMEFT, there are four leading-order BNV operators at dimension 6~\cite{Abbott:1980zj,Alonso:2014zka}, conserving the baryon minus lepton quantum number $\Delta(B-L)=0$. 
When these dimension-6 (dim-6) interactions are suppressed or even forbidden, dim-7 BNV operators provide the next relevant contributions~\cite{Lehman:2014jma,Liao:2016hru}. This is plausible because dim-7 operators carry different global baryon and lepton quantum numbers with $\Delta(B+L)=0$. 
Various ultraviolet(UV) models have been proposed in the literature that generate BNV interactions at dimension 7 at leading order~\cite{Li:2023cwy,Heeck:2026dmh}.

Within the SMEFT framework, the renormalization group (RG) evolution plays a crucial role in connecting physics at the new physics scale $\Lambda_{\tt NP}$ with the electroweak scale $\Lambda_{\tt EW}$.  
This evolution is essential for both fitting experimental data and interpreting constraints on new physics parameters.
The RG running effects have been extensively incorporated into phenomenological studies for the dim-6 SMEFT baryon number conserving sector.  
For the BNV sector, Ref.~\cite{Gisbert:2024sjw} has recently included the RG running effects to constrain Wilson coefficients (WCs) of dim-6 BNV operators involving the top quark. 
This approach is viable because the WCs, generated at the new physics scale, are defined in a flavor basis, whereas nucleon decays involve only light quarks in their mass eigenstates.
The rotation from the flavor to the mass basis inevitably mixes WCs involving the second or third generation fermions 
into those contributing directly to nucleon decays.
These flavor-to-mass rotation effects arise through two key steps: the RG evolution due to significant Yukawa mixing~\cite{Alonso:2014zka}, and the matching from the SMEFT onto the low-energy effective field theory (LEFT) at $\Lambda_{\tt EW}$~\cite{Jenkins:2017jig,Liao:2020zyx}. 

Given the dim-7 BNV operators associated with some high-scale UV theories, a comprehensive RG-improved analysis beyond a simple tree-level treatment becomes essential to fully constrain all relevant WCs at the high scale. 
In this work, we perform an RG-improved analysis for two-body nucleon decays into a light pseudoscalar meson and a lepton that are induced by dim-7 SMEFT BNV interactions, where the lepton can be either an electron $e^-$, a muon $\mu^-$, or a neutrino $\nu$. 
Employing the current experimental bounds summarized in \cref{tab:exp_bound}, we derive constraints on all 297 WCs of dim-7 SMEFT BNV operators, significantly surpassing a straightforward tree-level analysis.
We find that, in general, the indirect bounds on operators involving third- or double-second-generation fermions from RG mixing effects are weaker than those contributing to nucleon decays at tree level. Nevertheless, these constraints are still considerably stronger than the direct limits from processes involving heavy fermions. This highlights the importance of RG evolution when deriving bounds on new physics parameters from experimental data. 

\begin{table}[t]
\centering
\renewcommand{\arraystretch}{1.2}
\resizebox{0.4\linewidth}{!}{
\begin{tabular}{|l|r|}
\hline
\multirow{1}*{\quad~Mode}  
& \multirow{1}*{ $\makecell{ \Gamma^{-1}_{\rm exp.}(10^{30}\,\rm yr)~~}$}
\\\hline
$~p\to \nu_x \pi^{+} $ & 
$390$~\cite{Super-Kamiokande:2013rwg}~
\\
$~p\to \nu_x K^{+}~$ & $6.61\times10^3$~\cite{Mine:2016mxy}~
\\
\hline
$~n\to \nu_x \pi^0 $ & 
$1.4 \times10^3$~\cite{Super-Kamiokande:2025lxa}~
\\
$~n\to  \nu_x\eta $ & 
158~\cite{McGrew:1999nd}~
\\
$~n\to \nu_x K^0 $ & 
780~\cite{Super-Kamiokande:2025ibz}~
\\
$~n\to e^- K^+$ & 
$32$~\cite{Frejus:1991ben}~
\\
$~n\to \mu^- K^+$ & 
$57$~\cite{Frejus:1991ben}~
\\
\hline
\end{tabular} }
\caption{Current experimental bounds on two-body nucleon decay modes used in this work.} 
\label{tab:exp_bound}
\end{table}

%%%%%%%%%%%%%%%%%%%%%%%%
\section{Formalism for $\Delta(B+L)=0$ nucleon decays in SMEFT}
%%%%%%%%%%%%%%%%%%%%%%%%

%%%%%%%%%%%%%%%%%%%%%%%%
\subsection{SMEFT dim-7 BNV operators}
%%%%%%%%%%%%%%%%%%%%%%%%

In SMEFT, there are six dim-7 BNV operators~\cite{Lehman:2014jma,Liao:2016hru}.  In this study, we adopt the following conventions used in the {\tt D7RGESolver} code for solving the dim-7 RG equations~\cite{Liao:2019tep,Zhang:2023ndw,Liao:2025lxg},
\begin{subequations}
\label{eq:SMEFTdim7ope}
\begin{align}
\calO^{prst}_{\bar{L}dud\tilde{H}} 
=& \epsilon_{\alpha\beta\gamma}
(\overline{L_{p}} d^{\alpha}_{r}) 
(\overline{u^{\beta \C}_{s}} d^{\gamma}_{t}) \tilde{H},  
\\
\calO^{prst}_{\bar{L}dddH}
=\, &\epsilon_{\alpha\beta\gamma}
(\overline{L_{p}} d^{\alpha}_{r}) 
(\overline{d^{\beta\C}_{s}}d^{\gamma}_{t})H, 
\\
\calO^{prst}_{\bar{e}Qdd\tilde{H}}
=& \epsilon_{ij} \epsilon_{\alpha\beta\gamma}
( \overline{ e_{p}} Q^{i\alpha}_{r}) 
( \overline{d^{\beta \C}_{s}} d^{\gamma}_{t}) \tilde{H}^j,
\\
\calO^{prst}_{\bar{L}dQQ\tilde{H}} 
=& \epsilon_{ij}\epsilon_{\alpha\beta\gamma} 
(\overline{L_{p}} d^{\alpha}_{r} ) 
(\overline{Q^{\beta \C}_{s}} Q^{i\gamma}_{t}) \tilde{H}^j,
\\
\calO^{prst}_{\bar{L} QdDd}=& \epsilon_{\alpha\beta\gamma}
(\overline{L_{p}} \gamma_\mu Q^{\alpha}_{r})
(\overline{d^{\beta \C}_{\{s}} i D^\mu d^{\gamma}_{t\}} ),
\\
\calO^{prst}_{\bar{e}ddDd} =\, &
\epsilon_{\alpha \beta\gamma}
(\overline{e_{p}} \gamma_\mu d^{\alpha}_{\{r}) (\overline{d^{\beta \C}_{s}}i D^\mu d^{\gamma}_{t\}}),
\end{align}
\end{subequations}
where the curly brackets indicate symmetrization in all flavor indices of like-chirality fields, viz.,
$A_{\{y}B_{z\}} \equiv (1/2)(A_{y} B_{z} + A_{z}B_{y})$
and $A_{\{y}B_z C_{w\}} \equiv (1/6) [ A_{y}B_{z} C_w + \mbox{5 perms of }(y,z,w)]$.
The operator $\calO^{prst}_{\bar{L}dddH}$ has mixed flavor symmetries among the three right-handed down-type quarks:
$\calO^{prst}_{\bar{L}dddH}=-\calO^{prts}_{\bar{L}dddH}$ and $\calO^{prst}_{\bar{L}dddH}+\calO^{pstr}_{\bar{L}dddH}+\calO^{ptrs}_{\bar{L}dddH}=0$~\cite{Liao:2019tep}; the 
operators $\calO^{prst}_{\bar{e}Qdd\tilde{H}}$ and $\calO^{prst}_{\bar{L}QdDd}$ 
are respectively anti-symmetric and symmetric in the two down-type quarks: $\calO^{prst}_{\bar{e}Qdd\tilde{H}}=-\calO^{prts}_{\bar{e}Qdd\tilde{H}}$ and 
$\calO^{prst}_{\bar{L}QdDd}=\calO^{prts}_{\bar{L}QdDd}$; 
and the operator $\calO^{prst}_{\bar{e}ddDd}$ is totally symmetric among the three flavor indices $r,s,t$. 
The independent generation combinations chosen in our analysis are summarized in \cref{tab:dim7Ope_indp_flavor}.
In total, there are 297 independent operators after accounting for all generation indices (but not their hermitian conjugates). 

%%%%%%%%%%%%%%%
\begin{table}[h] 
\centering
\resizebox{0.95\linewidth}{!}{
\renewcommand{\arraystretch}{1.2}
\begin{tabular}{|c|c|l|}
\hline
~Classes~ 
&~Operators~
&~Independent generation combinations~
\\% 
\hline            
\multirow{6}{*}{$\psi^4 H$}
&$\calO^{prst}_{\bar{L}dud\tilde{H}}$
& $~(p,r,s,t)$
\\\cline{2-3}
& $\calO^{prst}_{\bar{L}dddH}$
&$\makecell[l]{~(p,1,1,2),~(p,2,1,2),~(p,3,1,2)\\
~(p,1,1,3),~(p,2,1,3),~(p,3,1,3)\\
~(p,2,2,3),~(p,3,2,3)}$
\\\cline{2-3}
& $\calO^{prst}_{\bar{e}Qdd\tilde{H}}$   
& $~(p,r,1,2),~(p,r,1,3),~(p,r,2,3)$
\\\cline{2-3}
& $ \calO^{prst}_{\bar{L}dQQ\tilde{H}}$ 
& $~(p,r,s,t)$
\\% 
\hline 
\multirow{4}{*}{$\psi^4D$}
& $\calO^{prst}_{\bar{L}QdDd}$ 
& $\makecell[l]{~(p,r,1,1),~(p,r,1,2),~(p,r,1,3)\\
~(p,r,2,2),~(p,r,2,3),~(p,r,3,3)}$
\\\cline{2-3}
&$\calO^{prst}_{\bar{e} ddDd}  $   
&$\makecell[l]{~(p,1,1,1),~(p,1,1,2),~(p,1,1,3)\\
~(p,1,2,2),~(p,1,2,3),~(p,1,3,3)\\
~(p,2,2,2),~(p,2,2,3),~(p,2,3,3)\\
~(p,3,3,3) }$
\\
\hline
\end{tabular}}
\caption{The independent generation combinations adopted in our analysis. The generation indices $p,\,r,\,s,\,t$ in the last column take 1,~2,~3. 
}
\label{tab:dim7Ope_indp_flavor}
\end{table}

%%%%%%%%%%%%%%%%%%%%%%%%
\subsection{LEFT operators}
%%%%%%%%%%%%%%%%%%%%%%%%

We focus on the two-body nucleon decays in \cref{tab:exp_bound} that can be induced by these dim-7 SMEFT operators at the leading order. We do not consider other processes such as triple-lepton modes mediated by additional SM electromagnetic or weak interactions~\cite{Chen:2025mjt}, as they yield weaker constraints.  
To calculate the decay widths, we employ the LEFT plus chiral perturbation theory (ChPT) framework developed in~\cite{Claudson:1981gh,Jenkins:2017jig,Liao:2025vlj}.
The relevant leading-order LEFT BNV operators are  dim-6 and dim-7 operators involving triple light quarks without being acted upon by a derivative.
They fall into the following three general independent structures and their chirality partners~\cite{Liao:2025vlj}, 
\begin{subequations}
\label{eq:LEFT3qO}
\begin{align}
{\cal O}_{a}^{yzw} =& \overline{\Psi_{a}}{\cal N}_{yzw}^{\tL\tL},
\\
{\cal O}_{b}^{yzw} =&\overline{\Psi_{b}}{\cal N}_{yzw}^{\tR\tL},
\\
{\cal O}_{c}^{yzw} =& i\overline{D_\mu\Psi_{c}}
{\cal N}_{yzw}^{\tL\tR,\mu},
\end{align}
\end{subequations}
where $\Psi$ is a charged lepton ($\ell=e,\,\mu$) or a neutrino ($\nu$) field in our consideration,
and the ${\cal N}$s represent triple-quark sectors in a definite irreducible representation (irrep) of the QCD chiral group $G_\chi\equiv\rm SU(3)_\tL\otimes SU(3)_\tR$, 
\begin{subequations}
\label{eq:Nyzw}
\begin{align}
{\cal N}_{yzw}^{\tL\tL} & =  q_{\tL, y}^\alpha (\overline{ q_{\tL, z}^{\beta \C} } q_{\tL, w}^\gamma )\epsilon_{\alpha \beta \gamma}\in \pmb{8}_\tL \otimes  \pmb{1}_\tR , 
\\
{\cal N}_{yzw}^{\tR\tL} & = q_{\tR, y}^\alpha (\overline{ q_{\tL, z}^{\beta \C} } q_{\tL,w}^\gamma)\epsilon_{\alpha \beta \gamma} \in 
\bar{\pmb{3}}_\tL \otimes \pmb{3}_\tR , 
\\
{\cal N}_{yzw}^{\tL\tR,\mu} & = q_{\tL,\{y}^\alpha (\overline{ q_{\tL, z\}}^{\beta \C} } \gamma^\mu q_{\tR,w}^\gamma)\epsilon_{\alpha \beta \gamma}
\in \pmb{6}_\tL \otimes \pmb{3}_\tR.
\end{align}
\end{subequations}
Here $\alpha,~\beta,~\gamma$ denote color and $y,~z,~w=1,~2,~3$ denote three light quarks with $q_{1,2,3}=u,~d,~s$. 
The curly brackets indicate symmetrization in the flavor indices of like-chirality fields as defined below \cref{eq:SMEFTdim7ope}.

Following the above conventions, we list in the second column of \cref{tab:matching_EW} the relevant leading-order LEFT operators involving at most one strange quark that contribute to the processes shown in \cref{tab:exp_bound}. The first column shows their corresponding chiral irreps. 
As shown in the table, nine dim-6 operators and four dim-7 operators (involving up to one strange quark $s$) can be generated by the dim-7 SMEFT BNV interactions through leading tree-level matching. In the next subsection, we describe in detail these matching relations presented in the third column of \cref{tab:matching_EW}.  

\begin{table*}[t]
\centering
\resizebox{1\linewidth}{!}{
\renewcommand{\arraystretch}{1.6}
\begin{tabular}{|c|c|c|c|}
\hline
~Chiral irreps.~
& LEFT operators
& Matching relations at $\Lambda_{\tt EW}$ in up (upper) and down (lower) flavor bases
&~Spurion fields at $\Lambda_{\chi}$~
\\
\hline
\multirow{8}*{\large\rotatebox[origin=c]{0}{
$\bar{\pmb{3}}_{\tL}\otimes \pmb{3}_{\tR}$}}& \multirow{2}*{$\calO_{ \bar{\ell} dds}^{\tR\tL,x}
= \overline{\ell_{\tL,x}}{\cal N}_{dds}^{\tR\tL}$ }
&$ 
C_{ \bar{\ell} dds}^{\tR\tL,x}=
-\frac{v}{\sqrt{2}}(V_{a1}V_{b2}-V_{a2}V_{b1})
\big[ C_{\bar{L}dQQ\tilde{H}}^{x1ab} +2(Y_d^{\dagger})_{wb}
C_{\bar{L}QdDd}^{xa1w} \big] $
&\multirow{2}*{$\mathcal{P}^{\tR\tL}_{dds}=1.32 C^{\tR\tL,x}_{\bar{\ell}dds}\overline{\ell_{\tL,x}}$}
\\\cline{3-3}%
& 
& $~C_{ \bar{\ell} dds}^{\tR\tL,x}=
-\frac{v}{\sqrt{2}} \big[C_{\bar{L}dQQ\tilde{H}}^{x112} 
-C_{\bar{L}dQQ\tilde{H}}^{x121}
+2(Y_d^{\dagger})_{w2}C_{\bar{L}QdDd}^{x11w}
-2(Y_d^{\dagger})_{w1}C_{\bar{L}QdDd}^{x21w}\big]~$ 
&
\\%
\cline{2-4}
&\multirow{4}*{$ \makecell{
\calO_{\bar{\nu}dud}^{\tR\tL,x}= 
\overline{\nu_{\tL,x}} {\cal N}_{dud}^{\tR\tL}\\
\calO_{\bar{\nu}dsu}^{\tR\tL,x}= \overline{\nu_{\tL,x}}{\cal N}_{dsu}^{\tR\tL} \\
\calO_{\bar{\nu}sud}^{\tR\tL,x}= \overline{\nu_{\tL,x}} {\cal N}_{sud}^{\tR\tL} }$  }
& \makecell{
$C_{\bar{\nu} dud}^{\tR\tL,x} =
-\frac{v}{\sqrt{2}}V_{a1}\big[C_{\bar{L}dQQ\tilde{H}}^{x11a}+2(Y_d^{\dagger})_{wa}C_{\bar{L}QdDd}^{x11w} \big]$
\\
$C_{\bar{\nu} dsu}^{\tR\tL,x} =
+\frac{v}{\sqrt{2}}V_{a2}\big[C_{\bar{L}dQQ\tilde{H}}^{x11a}+2(Y_d^{\dagger})_{wa}C_{\bar{L}QdDd}^{x11w} \big]$
\\
$C_{\bar{\nu} sud}^{\tR\tL,x} =
-\frac{v}{\sqrt{2}}V_{a1}\big[C_{\bar{L}dQQ\tilde{H}}^{x21a}+2(Y_d^{\dagger})_{wa}C_{\bar{L}QdDd}^{x12w}
\big]$ }
&\multirow{3}*[-1.2ex]{$\makecell[c]{\mathcal{P}^{\tR\tL}_{dud}=1.32 C^{\tR\tL,x}_{\bar{\nu}dud}\overline{\nu_{\tL,x}}
\\
\mathcal{P}^{\tR\tL}_{dsu}=1.32 C^{\tR\tL,x}_{\bar{\nu}dsu}\overline{\nu_{\tL,x}}
\\
\mathcal{P}^{\tR\tL}_{sud}=1.32 C^{\tR\tL,x}_{\bar{\nu}sud}\overline{\nu_{\tL,x}}
}$
}
\\\cline{3-3}
&
& \makecell{
$C_{\bar{\nu} dud}^{\tR\tL,x} =
-\frac{v}{\sqrt{2}}V_{1a}^*\big[ C_{\bar{L}dQQ\tilde{H}}^{x1a1}+2(Y_d^{\dagger})_{w1}C_{\bar{L}QdDd}^{xa1w} \big]$
\\
$C_{\bar{\nu} dsu}^{\tR\tL,x} =
+\frac{v}{\sqrt{2}}V_{1a}^* \big[C_{\bar{L}dQQ\tilde{H}}^{x1a2}+2(Y_d^{\dagger})_{w2}C_{\bar{L}QdDd}^{xa1w} \big]$
\\
$C_{\bar{\nu} sud}^{\tR\tL,x} =
-\frac{v}{\sqrt{2}}V_{1a}^* \big[C_{\bar{L}dQQ\tilde{H}}^{x2a1}+2(Y_d^{\dagger})_{w1}C_{\bar{L}QdDd}^{xa2w} \big]$ }
&
\\%
\hline
\multirow{2}*{\large\rotatebox[origin=c]{0}{
$\pmb{3}_{\tL}\otimes \bar{\pmb{3}}_{\tR}$}}& \multirow{2}*{$\calO_{\bar{\ell}dds}^{\tL\tR,x} = \overline{\ell_{\tR,x}} {\cal N}_{dds}^{\tL\tR} $ }
& \makecell{ $C_{\bar{\ell} dds}^{\tL\tR,x} = 
-\sqrt{2}vV_{a1}C_{\bar{e}Qdd\tilde{H}}^{xa12}$}
&\multirow{2}*{$\mathcal{P}^{\tL\tR}_{dds}=1.32 C^{\tL\tR,x}_{\bar{\ell}dds}\overline{\ell_{\tR,x}}$}
\\\cline{3-3}
& 
& \makecell{ $C_{\bar{\ell} dds}^{\tL\tR,x} = 
-\sqrt{2}v C_{\bar{e}Qdd\tilde{H}}^{x112}~~~~\,$}
&
\\%
\hline
\multirow{2}*{\large\rotatebox[origin=c]{0}{
$\pmb{1}_{\tL}\otimes \pmb{8}_{\tR}$}}& $ \makecell{\calO_{\bar{\nu} dud}^{\tR\tR,x}= 
\overline{\nu_{\tL,x}} {\cal N}_{dud}^{\tR\tR} \\
\calO_{\bar{\nu} dsu}^{\tR\tR,x}= 
\overline{\nu_{\tL,x}} {\cal N}_{dsu}^{\tR\tR} \\
\calO_{\bar{\nu} sud}^{\tR\tR,x}= 
\overline{\nu_{\tL,x}} {\cal N}_{sud}^{\tR\tR} }$
& \makecell{
$C_{\bar{\nu} dud}^{\tR\tR,x} = 
+\frac{v}{\sqrt{2}}\big[C_{\bar{L}dud\tilde{H}}^{x111}+(Y_u)_{w1}C_{\bar{L}QdDd}^{xw11}\big] $
\\
$C_{\bar{\nu} dsu}^{\tR\tR,x} = 
-\frac{v}{\sqrt{2}}\big[C_{\bar{L}dud\tilde{H}}^{x112}+(Y_u)_{w1}C_{\bar{L}QdDd}^{xw12}\big] $
\\
$C_{\bar{\nu} sud}^{\tR\tR,x} = 
+\frac{v}{\sqrt{2}}\big[C_{\bar{L}dud\tilde{H}}^{x211}+(Y_u)_{w1}C_{\bar{L}QdDd}^{xw12}\big] $}
&\multirow{2}*[1.2ex]{$\makecell[c]{\mathcal{P}^{\tR\tR}_{dud}=1.32 C^{\tR\tR,x}_{\bar{\nu}dud}\overline{\nu_{\tL,x}}
\\
\mathcal{P}^{\tR\tR}_{dsu}=1.32 C^{\tR\tR,x}_{\bar{\nu}dsu}\overline{\nu_{\tL,x}}
\\
\mathcal{P}^{\tR\tR}_{sud}=1.32 C^{\tR\tR,x}_{\bar{\nu}sud}\overline{\nu_{\tL,x}}
\\
\mathcal{P}^{\tR\tR}_{dds}=1.32 C^{\tR\tR,x}_{\bar{\ell}dds}\overline{\ell_{\tL,x}}
}$
}
\\%
\cline{2-3}
& $\calO_{ \bar{\ell}dds}^{\tR\tR,x} = \overline{\ell_{\tL,x}} {\cal N}_{dds}^{\tR\tR}$
& $C_{ \bar{\ell} dds}^{\tR\tR,x} =
\frac{v}{\sqrt{2}}\big[2C_{LdddH}^{x112}
-(Y_d)_{w2}C_{\bar{L}QdDd}^{xw11}
+(Y_d)_{w1}C_{\bar{L}QdDd}^{xw12}\big]$
&
\\%
\hline
\multirow{6}*{\large\rotatebox[origin=c]{0}{
$\pmb{3}_{\tL}\otimes \pmb{6}_{\tR}$}}
&\multirow{2}*{$\makecell[l]{
\calO_{\partial\bar{\ell}ddd}^{\tR\tL} 
= i\overline{D_\mu\ell_{\tL,x}}{\cal N}_{ddd}^{\tR\tL,\mu}\\
\calO_{\partial\bar{\ell}dsd}^{\tR\tL} 
= i\overline{D_\mu\ell_{\tL,x}}{\cal N}_{dsd}^{\tR\tL,\mu}\\
\calO_{\partial\bar{\ell}sdd}^{\tR\tL} 
= i\overline{D_\mu\ell_{\tL,x}}{\cal N}_{sdd}^{\tR\tL,\mu}\\
\calO_{\partial\bar{\ell}dds}^{\tR\tL} 
= i\overline{D_\mu\ell_{\tL,x}}{\cal N}_{dds}^{\tR\tL,\mu}}$}
& $\makecell{ C_{\partial\bar{\ell} ddd}^{\tR\tL,x} =
-V_{w1}C_{\bar{L}QdDd}^{xw11},~
C_{\partial\bar{\ell} dsd}^{\tR\tL,x} =
-V_{w1}C_{\bar{L}QdDd}^{xw12} \\
C_{\partial\bar{\ell} sdd}^{\tR\tL,x} =
-V_{w1}C_{\bar{L}QdDd}^{xw12},~
C_{\partial\bar{\ell} dds}^{\tR\tL,x} =
-V_{w2}C_{\bar{L}QdDd}^{xw11}}$
&\multirow{6}*[-1ex]{$\makecell[c]{
\mathcal{P}^{\tR\tL,\mu}_{ddd}=0.91 C^{\tR\tL,x}_{\partial\bar{\ell}ddd}i\overline{D^\mu\ell_{\tL,x}}
\\
\mathcal{P}^{\tR\tL,\mu}_{dsd}=0.91 C^{\tR\tL,x}_{\partial\bar{\ell}dsd}i\overline{D^\mu\ell_{\tL,x}}
\\
\mathcal{P}^{\tR\tL,\mu}_{sdd}=0.91 C^{\tR\tL,x}_{\partial\bar{\ell}sdd}i\overline{D^\mu\ell_{\tL,x}}
\\
\mathcal{P}^{\tR\tL,\mu}_{dds}=0.91 C^{\tR\tL,x}_{\partial\bar{\ell}dds}i\overline{D^\mu\ell_{\tL,x}}
\\
\mathcal{P}^{\tR\tL,\mu}_{ddu}=0.91 C^{\tR\tL,x}_{\partial\bar{\nu}ddu}i\overline{\partial^\mu\nu_{\tL,x}}
\\
\mathcal{P}^{\tR\tL,\mu}_{dsu}=0.91 C^{\tR\tL,x}_{\partial\bar{\nu}dsu}i\overline{\partial^\mu\nu_{\tL,x}}
\\
\mathcal{P}^{\tR\tL,\mu}_{sdu}=0.91 C^{\tR\tL,x}_{\partial\bar{\nu}sdu}i\overline{\partial^\mu\nu_{\tL,x}}
}$
}
\\\cline{3-3}%
& 
& $\makecell{ C_{\partial\bar{\ell} ddd}^{\tR\tL,x} = -C_{\bar{L}QdDd}^{x111},~
C_{\partial\bar{\ell} dsd}^{\tR\tL,x} =
-C_{\bar{L}QdDd}^{x112} \\
C_{\partial\bar{\ell} sdd}^{\tR\tL,x} =
-C_{\bar{L}QdDd}^{x112},~
C_{\partial\bar{\ell} dds}^{\tR\tL,x} =
-C_{\bar{L}QdDd}^{x211} }$
&
\\\cline{2-3}%
& \multirow{4}{*}{
$\makecell[l]{
\calO_{\partial\bar{\nu} dd u}^{\tR\tL,x}=
i\overline{\partial_\mu\nu_{\tL,x}}{\cal N}_{ddu}^{\tR\tL,\mu} \\ 
\calO_{\partial\bar{\nu} ds u}^{\tR\tL,x}=
i\overline{\partial_\mu\nu_{\tL,x}}{\cal N}_{dsu}^{\tR\tL,\mu} \\ 
\calO_{\partial\bar{\nu} sdu}^{\tR\tL,x}=
i\overline{\partial_\mu\nu_{\tL,x}}{\cal N}_{sdu}^{\tR\tL,\mu} }$ }
&\multirow{2}{*}{$ C_{\partial \bar{\nu} ddu}^{\tR\tL,x} =
-C_{\bar{L}QdDd}^{x111},~
C_{ \partial\bar{\nu} dsu}^{\tR\tL,x} =
-C_{\bar{L}QdDd}^{x112},~ 
C_{ \partial\bar{\nu} sdu}^{\tR\tL,x} =
-C_{\bar{L}QdDd}^{x112}$}
& 
\\
& & &
\\\cline{3-3}%
&
&\multirow{2}{*}{ $ C_{\partial\bar{\nu} ddu}^{\tR\tL,x} =
-V^\star_{1w}C_{\bar{L}QdDd}^{xw11},~
C_{\partial\bar{\nu} dsu}^{\tR\tL,x} =
-V^\star_{1w}C_{\bar{L}QdDd}^{xw12},~
C_{\partial\bar{\nu} sdu}^{\tR\tL,x} =
-V^\star_{1w}C_{\bar{L}QdDd}^{xw12} $}
&
\\
& & &
\\% 
\hline
\end{tabular} }
\caption{The LEFT BNV operator basis used in our analysis and the corresponding tree-level matching relations with the dim-7 SMEFT BNV interactions, shown for both the up- and down-quark flavor bases.}
\label{tab:matching_EW}
\end{table*}

%%%%%%%%%%%%%%%%%%%%%%%%
\subsection{Matching onto the LEFT}
%%%%%%%%%%%%%%%%%%%%%%%%

Next, we consider tree-level matching of the SMEFT interactions onto the LEFT interactions at the electroweak scale $\Lambda_{\tt EW}$. This matching depends on the choice of a quark flavor basis. Following Refs.~\cite{Aebischer:2020dsw,Liao:2024xel,Liao:2025lxg}, we consider both scenarios: the up-quark and down-quark flavor basis. 
We choose charged leptons in their mass eigenstates and neglect tiny neutrino mass effects. 
In the up-quark flavor basis, both the left- and right-handed up-type quark fields, as well as the right-handed down-type quark fields, are already mass eigenstates. In this case, the weak eigenstates $d'_{\tL}$ and the mass eigenstates $d_{\tL}$ of the left-handed down-type quarks are related by the Cabibbo-Kobayashi-Maskawa (CKM) matrix $V$~\cite{Cabibbo:1963yz, Kobayashi:1973fv} through $d'_{\tL}=V d_{\tL}$.
Conversely, in the down-quark flavor basis, the weak and mass eigenstates of the left-handed up-type quarks are related by $u_\tL'=V^\dagger u_\tL$.

By replacing the SM Higgs field with its vacuum expectation value (VEV) $v=(\sqrt{2} G_F)^{-1/2}\simeq 246$ GeV, and taking into account the two flavor basis choices, we present the matching results between the SMEFT and LEFT WCs in the third column of \cref{tab:matching_EW}. For each cell, 
the top (bottom) row corresponds to the up (down)-quark flavor basis. The results are shown in a single row when the matching is basis-independent. 
For the four dim-7 SMEFT operators involving a Higgs field, our matching results agree with those presented in~\cite{Liao:2020zyx,Ma:2025mjy} after accounting for differences in operator conventions and flavor indices.
For the operator $\calO_{\bar{L} QdDd}$ where the derivative acts on the quark field, we transfer the derivative to the lepton field using Fierz identities, integration by parts, and equations of motion of quark fields. The resulting expression is
\begin{align}
\calO^{prst}_{\bar{L} QdDd }=\, & 
\frac{1}{2}\Big\{\big[
(Y_{\rm u})_{rw}\calO_{\bar{L}dud\tilde{H}}^{pswt}
-(Y_{\rm d})_{rw}\calO_{\bar{L}dddH}^{pstw}
\nonumber\\
&+2 (Y_{\rm d}^{\dagger})_{sw}\calO_{\bar{L}dQQ \tilde{H}}^{ptrw}\big] + s\leftrightarrow t\Big\}
\nonumber\\
&-\epsilon_{\alpha\beta\gamma}
(i \overline{D^\mu L_{p}}  d^{\alpha}_{\{s})
(\overline{d^{\beta \C}_{t\}}} \gamma_\mu Q^{\gamma}_{r}).
\label{eq:OLQdDd_res}
\end{align}
The details of this transformation are provided in Appendix \ref{app:convert_LQddD}.
The first two lines on the right-hand side of \cref{eq:OLQdDd_res} modify the WCs of three operators involving the Higgs, while the last term yields LEFT operators in the chiral irrep $\pmb{3}_{\tL}\otimes \pmb{6}_{\tR}$. The matching contribution from $\calO^{prst}_{\bar{L} QdDd }$ is characterized by $C_{\bar{L} QdDd }$ in \cref{tab:matching_EW}.
The second derivative-type SMEFT operator $\calO^{prst}_{\bar{e}ddDd}$ can only be matched onto LEFT operators with a derivative acting on quark fields.
Since the chiral realization of this operator is not available in the literature, and we leave their direct contribution to the nucleon decay processes for the future study. Instead, we derive constraints on their WCs through the RG mixing effects. 

%%%%%%%%%%%%%%%%%%%%%%%%
\subsection{Chiral matching}
%%%%%%%%%%%%%%%%%%%%%%%%

We now collect the chiral matching results for the LEFT interactions in \cref{tab:matching_EW}.
Within the ChPT framework, the quark-level interactions are mapped onto interactions among octet baryons, octet pseudoscalar mesons, and non-QCD fields such as leptons and photons.  
This chiral realization can be carried out systematically using the spurion field method, where the WCs and the non-QCD fields are collected in the spurion fields ${\cal P}$s that transform appropriately under the chiral group $G_\chi$. In the last column of \cref{tab:matching_EW}, we list the corresponding spurion field for each triple-quark operator. Since chiral matching is performed at the chiral symmetry breaking scale $\Lambda_\chi\sim 1.2\,\rm GeV$, we have included the effects of QCD RG evolution from $\Lambda_{\tt EW}$ down to $\Lambda_\chi$ in the spurion fields ${\cal P}$s, as reflected in the numerical factors 1.32 and 0.91~\cite{Liao:2025vlj}. 

By organizing the ${\cal N}$s and the corresponding spurion fields ${\cal P}$s in the irreps ${\bf 8}_{\tL(\tR)}\otimes {\bf 1}_{\tR(\tL)}$ and $\bar{\pmb{3}}_{\tL(\tR)} \otimes \pmb{3}_{\tR(\tL)}$ in matrix form, see \cref{eq:3qpart}, 
the $\Delta B=1$ Lagrangian involving the three structures in \cref{eq:Nyzw} can be compactly expressed as~\cite{Liao:2025vlj} 
\begin{align}
{\cal L}_{q^3}^{\slashed{B}} & = 
{\rm Tr} \big[  
  {\cal P}_{\pmb{8}_\tL \otimes \pmb{1}_\tR }
  {\cal N}_{\pmb{8}_\tL \otimes \pmb{1}_\tR } 
+ {\cal P}_{ \pmb{1}_\tL \otimes \pmb{8}_\tR }  
  {\cal N}_{  \pmb{1}_\tL \otimes \pmb{8}_\tR }
  \big]  
\nonumber 
\\
& + {\rm Tr} \big[ 
  {\cal P}_{\pmb{3}_\tL \otimes \bar{\pmb{3}}_\tR }
  {\cal N}_{\bar{\pmb{3}}_\tL \otimes \pmb{3}_\tR } 
+ {\cal P}_{\bar{\pmb{3}}_\tL \otimes \pmb{3}_\tR }
  {\cal N}_{\pmb{3}_\tL \otimes \bar{\pmb{3}}_\tR }
 \big] 
\nonumber 
\\
& + \big[
{\cal P}_{yzw}^{\tL\tR,\mu}
{\cal N}_{yzw,\mu}^{\tL\tR}
+ {\cal P}_{yzw}^{\tR\tL,\mu}
{\cal N}_{yzw,\mu}^{\tR\tL}
\big],
\label{eq:q3LEFT}
\end{align}
where summation over indices $y,z,w$ is implied. 
Denoting the matrices for the octet baryon and pseudoscalar meson fields as $B(x)$ and $\Pi(x)$, respectively, the leading-order chiral matching takes the following form~\cite{Liao:2025vlj}, 
\begin{align}
{\cal L}_{B}^{\slashed{B}} =
& c_1 {\rm Tr}\big[ 
{\cal P}_{  \bar{\pmb{3}}_\tL \otimes \pmb{3}_\tR} \xi B_\tL \xi -
{\cal P}_{\pmb{3}_\tL \otimes \bar{\pmb{3}}_\tR} \xi^\dagger B_\tR \xi^\dagger 
 \big]
 \nonumber\\
& + c_2 {\rm Tr}\big[ 
{\cal P}_{\pmb{8}_\tL \otimes \pmb{1}_\tR}\xi B_\tL \xi^\dagger
- {\cal P}_{ \pmb{1}_\tL \otimes  \pmb{8}_\tR} \xi^\dagger B_\tR \xi
\big] 
\nonumber
\\
& + {c_3 \over \Lambda_\chi} \big[ 
{\cal P}_{yzi}^{\tL\tR,\mu}
{\Gamma}_{\mu\nu}^{\tt L} 
(\xi i D^\nu B_\tL \xi)_{yj}
\Sigma_{zk} \epsilon_{ijk}
\nonumber
\\
&- {\cal P}_{yzi}^{\tR\tL,\mu}
{\Gamma}_{\mu\nu}^{\tt R}
(\xi^\dagger i D^\nu B_\tR  \xi^\dagger)_{yj}
\Sigma^*_{kz} \epsilon_{ijk} \big]
+\text{H.c.},
\label{eq:LBlM}
\end{align}%
where $\Sigma = \xi^2 ={\rm exp}[{2 i \Pi/f_{\pi}}]$, 
with $f_{\pi}$ being the pion decay constant in the chiral limit. Numerically, we take $f_{\pi}=130.41(20)\,\rm MeV$~\cite{ParticleDataGroup:2024cfk}.
The projectors ${\Gamma}_{\mu\nu}^{\tt L,R} \equiv (g_{\mu\nu} - {1\over 4}\gamma_\mu\gamma_\nu)P_{\tt L,R}$.
The coefficients $c_{1,2,3}$ are low-energy constants (LECs), where
$c_1=-0.01257(111)\,{\rm GeV}^3$ and
$c_2=0.01269(107)\,{\rm GeV}^3$ are taken from recent lattice QCD calculations~\cite{Yoo:2021gql}, 
while $c_3 \approx 0.011\,{\rm GeV}^3$ is estimated using naive dimensional analysis~\cite{Liao:2025vlj}. 
We use the central values as our numerical inputs and neglect the uncertainties, since they affect the results only at a subleading level. 

By substituting the spurion fields from \cref{tab:matching_EW} into \cref{eq:LBlM} and expanding the pseudoscalar exponential up to the first order in the meson fields, we obtain the baryon-lepton ($Bl$) and baryon-lepton-meson ($BlM$) vertices that mediate nucleon decays. The leading-order Feynman diagrams for the two-body decay processes are shown in \cref{fig:Feyndiagram}. 
The relevant vertices and decay widths are summarized in Appendix \ref{app:chiral_matching}.

\begin{figure}[t]
\centering
\begin{tikzpicture}[mystyle,scale=1]
\begin{scope}
\draw[f] (0, 0)node[left]{$\N$} -- (1.5,0);
\draw[f] (1.5, 0) -- (3,0) node[right]{$l$};
\draw[snar, black] (1.5,0) -- (2.5,1.2) node[right,yshift = 2 pt]{$M$};
\filldraw [cyan] (1.5,0) circle (3pt);
\end{scope}
\end{tikzpicture}
\quad 
\begin{tikzpicture}[mystyle,scale=1]
\begin{scope}
\draw[f] (0, 0)node[left]{$\N$} -- (1.5,0);
\draw[f] (1.5, 0) -- (3,0) node[midway,xshift = 5 pt,yshift = 6 pt]{$B$};
\draw[snar, black] (1.5,0) -- (2.5,1.2) node[right,yshift = 2 pt]{$M$};
\draw[f] (3.0, 0) -- (4.5,0) node[right]{$l$};
\filldraw [black] (1.38,-0.12) rectangle (1.62,0.12);
\filldraw [cyan] (3,0) circle (3pt);
\end{scope}
\end{tikzpicture}
\caption{ Diagrams for the contact (left) and non-contact (right) contributions to two-body nucleon decays into a meson and a lepton. The cyan blob (black square) represents the insertion of a BNV (usual) chiral vertex.}
\label{fig:Feyndiagram}
\end{figure}

%%%%%%%%%%%%%%%%%%%%%%%%
\section{RG-improved analysis of nucleon decays}
%%%%%%%%%%%%%%%%%%%%%%%%

Having derived the matching relations and nucleon decay widths in the previous section, we now solve the SMEFT RG equations to establish the constraints on all relevant WCs at higher energy scales.

%%%%%%%%%%%%%%%%%%%%%%%%
\subsection{Solving the RG equations}
%%%%%%%%%%%%%%%%%%%%%%%%

The complete RG equations for the dim-7 operators in \cref{eq:SMEFTdim7ope} due to SM corrections have been worked out in~\cite{Liao:2016hru,Zhang:2023ndw}, and implemented into the numerical code {\tt D7RGESolver}~\cite{Liao:2025lxg}. 
Since the solutions to these RGEs depend on the specific form of the Yukawa matrices, 
we work in both the up- and down-quark flavor bases to maintain consistency with the matching conditions. 
In the up-quark flavor basis,
the up-type-quark Yukawa matrix is diagonal at the electroweak scale $\Lambda_{\tt EW}$, while the down-type Yukawa matrix incorporates the CKM matrix at $\Lambda_{\tt EW}$,
\begin{align}
\label{eq:up}
Y_u=\frac{\sqrt{2}}{v} {{M}_u},\quad
Y_d=\frac{\sqrt{2}}{v} V {{M}_d},
\end{align}
where the up- and down-type quark mass matrices are diagonal, with ${M}_u = \text{diag}(m_u,~m_c,~m_t)$ and ${M}_d = \text{diag}(m_d,~m_s,~m_b)$, respectively. In the down-quark flavor basis, the quark Yukawa matrices at $\Lambda_{\tt EW}$ are
\begin{align}
\label{eq:down}
Y_u=\frac{\sqrt{2}}{v} V^{\dagger}{{M}_u},\quad Y_d=\frac{\sqrt{2}}{v}{{M}_d}.
\end{align}
In both bases, the charged lepton Yukawa matrix $Y_l$ is taken to be diagonal at $\Lambda_{\tt EW}$
\begin{align}
Y_l=\frac{\sqrt{2}}{v} {{M}_e}\;,
\end{align}
where ${M}_e = \text{diag}(m_e, m_\mu, m_\tau)$ is the charged lepton mass matrix.
Note that $Y_u$ ($Y_d$) in the up (down)-quark flavor basis is defined to be diagonal only at $\Lambda_{\tt EW}$.
At other energy scales, these matrices generally become non-diagonal because their RGEs involve non-diagonal anomalous dimension matrices from the Yukawa interactions~\cite{Arason:1991ic}. In contrast, the lepton Yukawa matrix $Y_l$ remains diagonal under RG evolution. 

Assuming these BNV interactions are generated at a new physics scale $\Lambda_{\tt NP}=10^8\,\rm GeV$, we evaluate the RG running effects. 
Note that our qualitative conclusions remain robust under variations in the new physics scale, with WC constraints exhibiting at most an $\calO(1)$ factor of change. 
For example, consider the WC $C_{\bar{L}dud\tilde H}^{p111}$ at two scales: $\Lambda_{\tt NP}=10^8\,\rm GeV$ and $\Lambda_{\tt NP}'=10^5\,\rm GeV$. After RG evolution to $\Lambda_{\tt EW}$, the resulting outputs are
\begin{subequations}
\begin{align}
C_{\bar{L}dud\tilde H}^{p111}(\Lambda_{\tt EW})&\approx 1.33 C_{\bar{L}dud\tilde H}^{p111}(\Lambda_{\tt NP}),
\\
C_{\bar{L}dud\tilde H}^{p111}(\Lambda_{\tt EW})&\approx 1.18 C_{\bar{L}dud\tilde H}^{p111}(\Lambda_{\tt NP}').
\end{align}
\end{subequations}
This indicates that the constraint on $C_{\bar{L}dud\tilde H}^{p111}$ differs by a factor of 1.13 between the two scales. Consequently, we restrict our analysis to $\Lambda_{\tt NP}=10^8\,\rm GeV$ in what follows. 

%%%%%%%%%%%%%%%%%%%%%%%%
\subsection{Constraints at a higher new physics scale}
%%%%%%%%%%%%%%%%%%%%%%%%

We analyze one WC at the scale $\Lambda_{\tt NP}$ at a time,
following the generation combination scheme presented in~\cref{tab:dim7Ope_indp_flavor}.\footnote{One can examine the scenario involving two WCs simultaneously as we did previously in~\cite{Liao:2025lxg}. In that case, the resulting bounds form a band-like or elliptical shape.}
By evolving it down to $\Lambda_{\tt EW}$ using {\tt D7RGESolver},
we match it onto the LEFT at $\Lambda_{\tt EW}$ according to \cref{tab:matching_EW}, and then compute the inverse decay widths for all relevant decay modes based on \cref{eq:LB2lM} and \cref{tab:matching_EW,tab:B2lM_vertex}.
This procedure yields the inverse decay widths as a function of the input SMEFT WC defined at $\Lambda_{\tt NP}$. 
To establish lower constraints on the WC, we equate the derived expressions with the current experimental bounds in \cref{tab:exp_bound} and select the most stringent limit from all available decay channels as our final constraint. See Appendix \ref{app:example} for an illustrative example of the procedure.
For operators involving flavor symmetries,
we adopt the convention that their corresponding WCs are generated with equal magnitude at the new physics scale. For instance, we take 
$C^{pr21}_{\bar{e}Qdd\tilde{H}}
=-C^{pr12}_{\bar{e}Qdd\tilde{H}}$,  
$C^{pr21}_{\bar{L}QdDd}
=C^{pr12}_{\bar{L}QdDd}$, 
and $C^{p211}_{\bar{e}ddDd}
=C^{p121}_{\bar{e}ddDd}
=C^{p112}_{\bar{e}ddDd}$, etc.
For the operator $\calO_{\bar{L}dddH}^{p123}$, we use the convention 
$C_{\bar{L}dddH}^{p123}=
C_{\bar{L}dddH}^{p213}
-C_{\bar{L}dddH}^{p312}$, 
as dictated by the mixed flavor symmetry properties of the operator.

\begin{figure*}[t]
\centering
\includegraphics[width=0.31\linewidth]{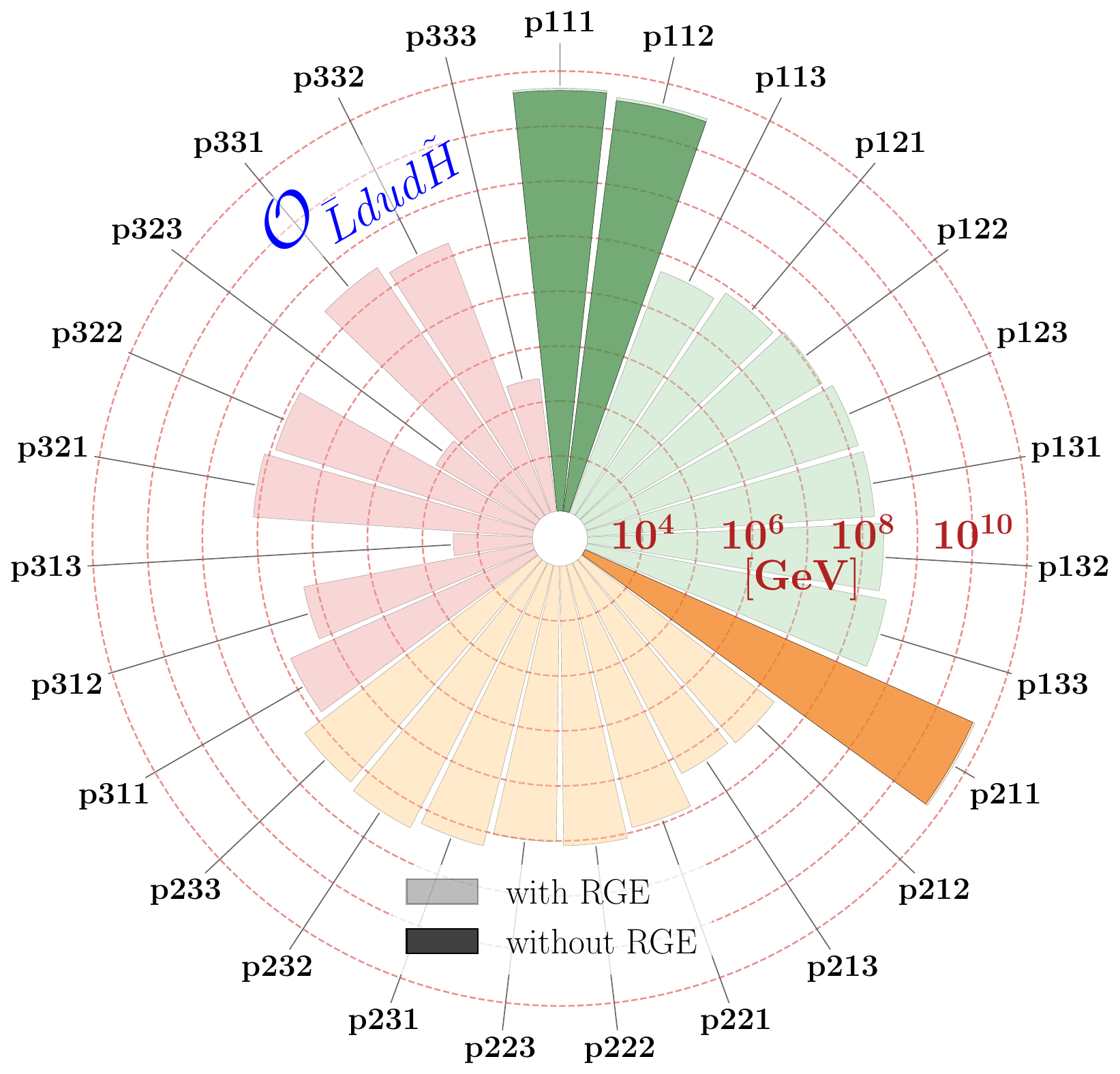}~
\includegraphics[width=0.31\linewidth]{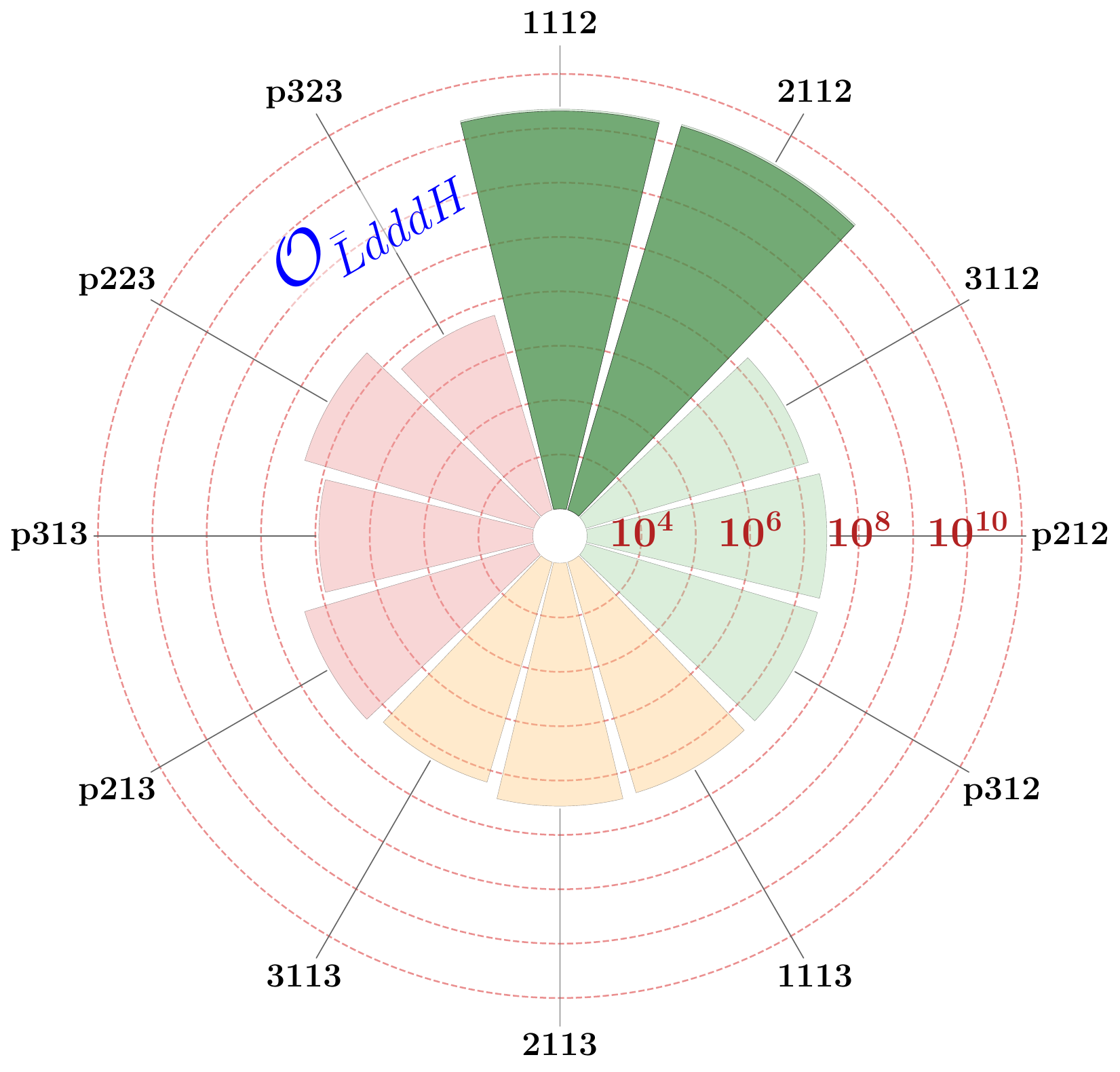}~
\includegraphics[width=0.31\linewidth]{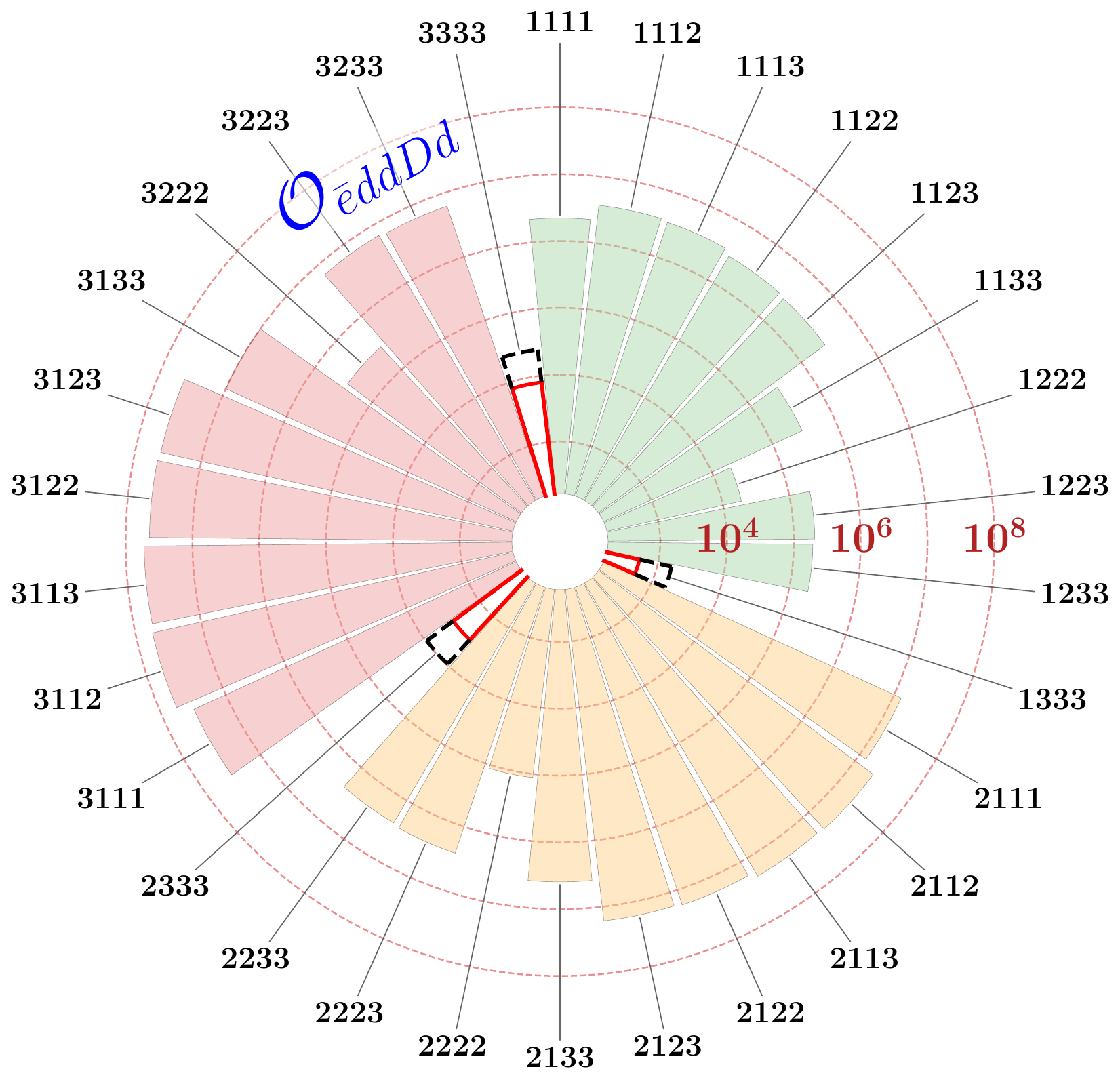}
\\[2pt]
\includegraphics[width=0.31\linewidth]{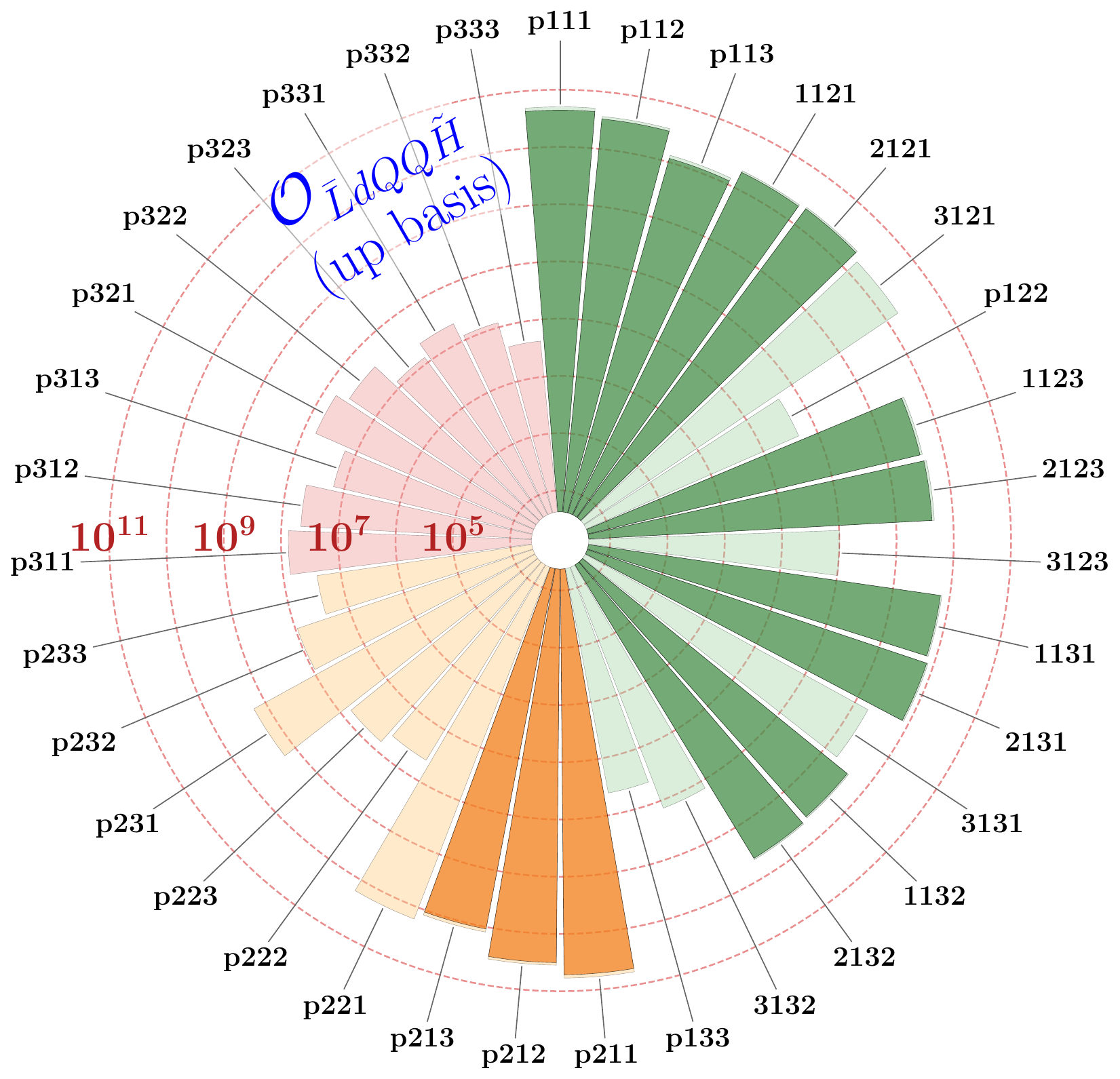}~
\includegraphics[width=0.31\linewidth]{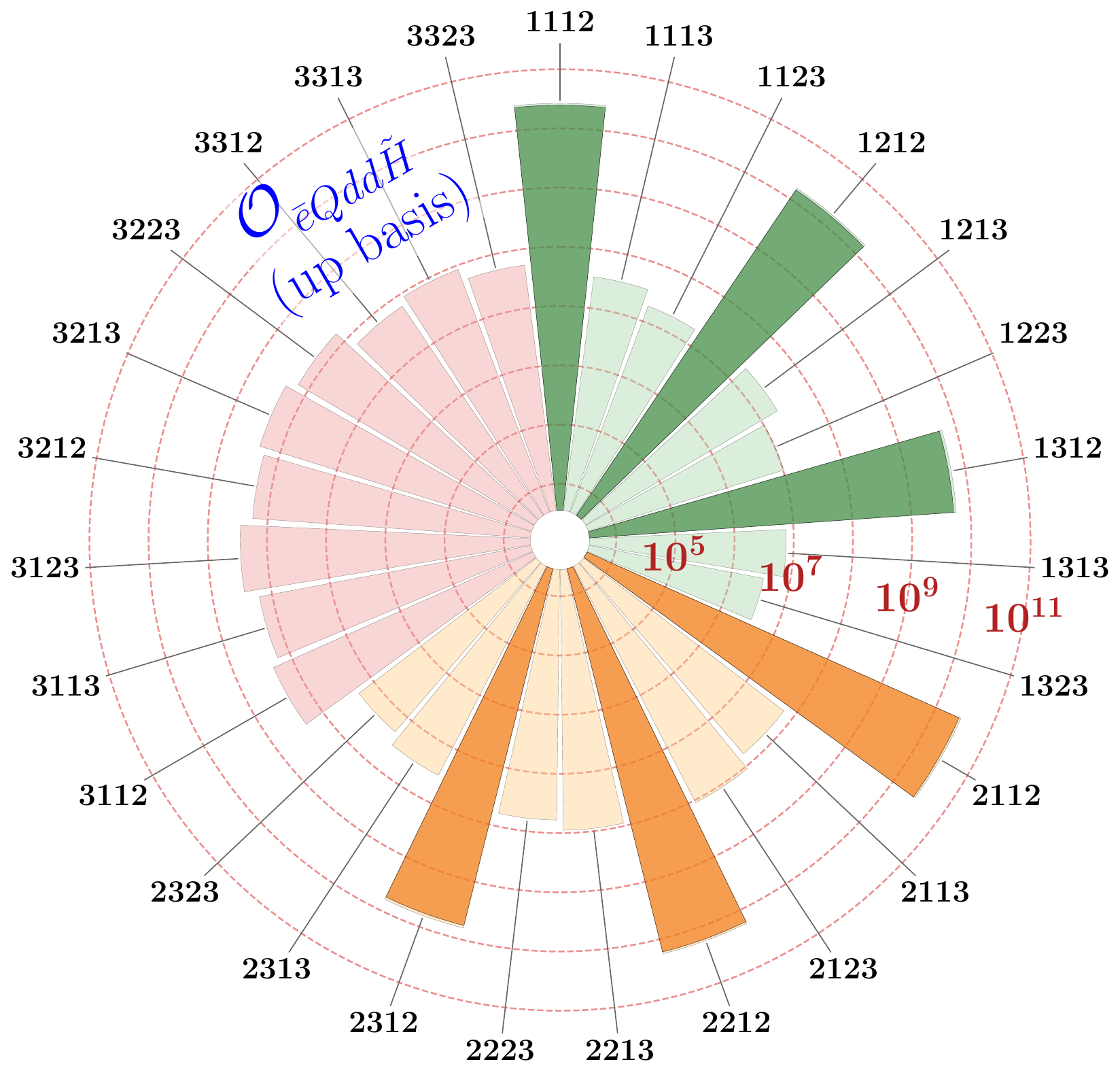}~
\includegraphics[width=0.31\linewidth]{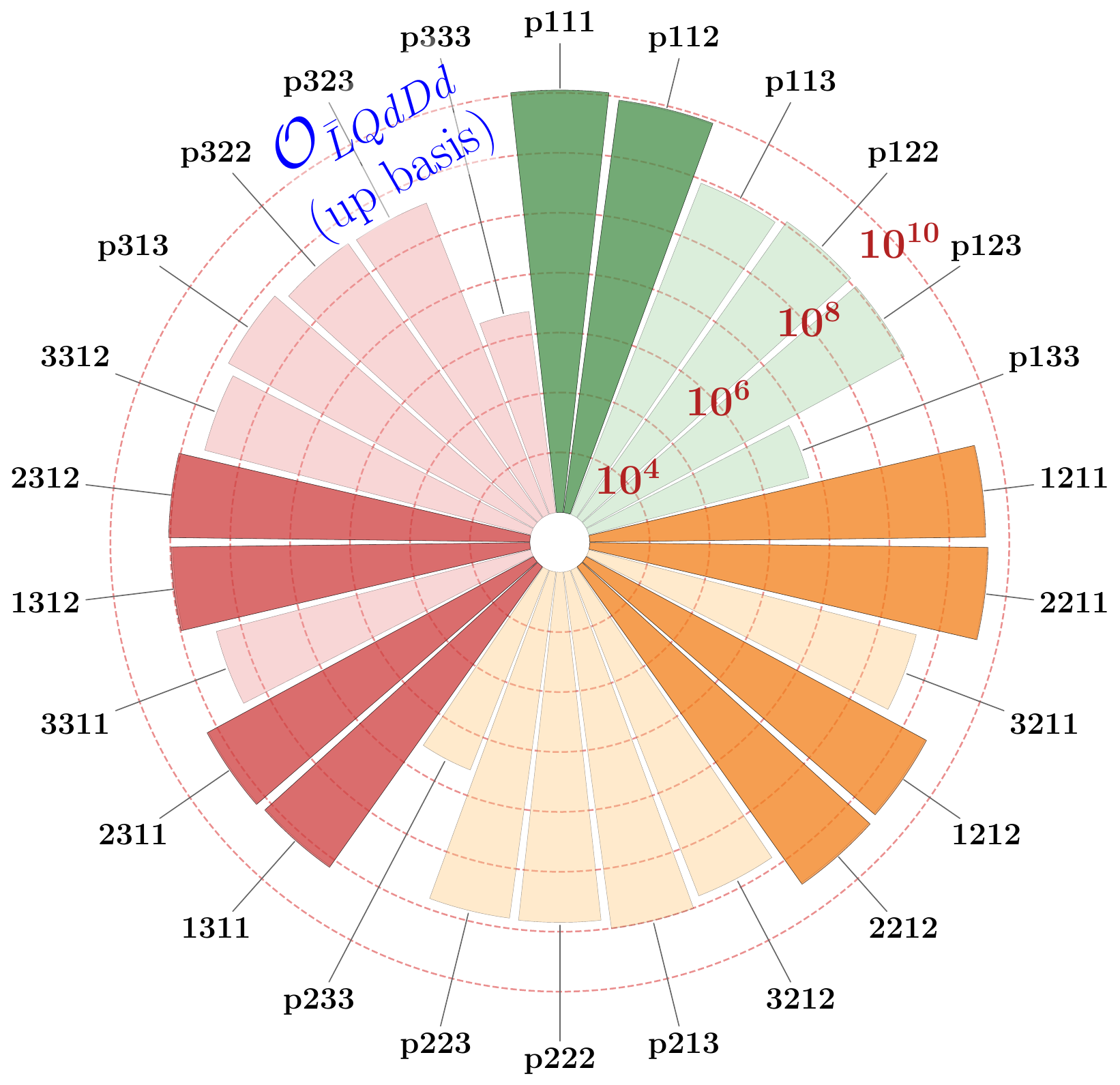}
\\[2pt]
\includegraphics[width=0.31\linewidth]{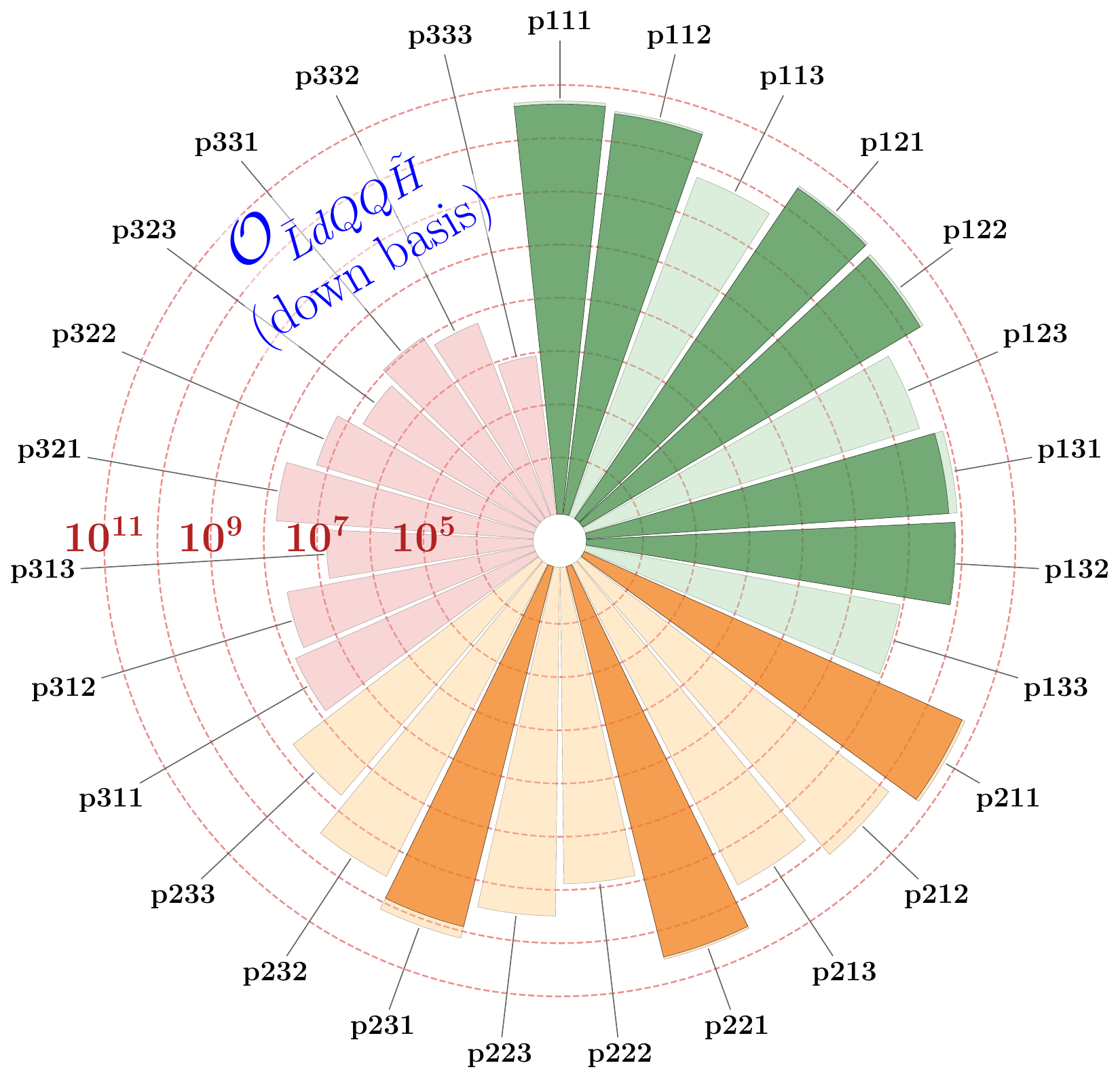}~
\includegraphics[width=0.31\linewidth]{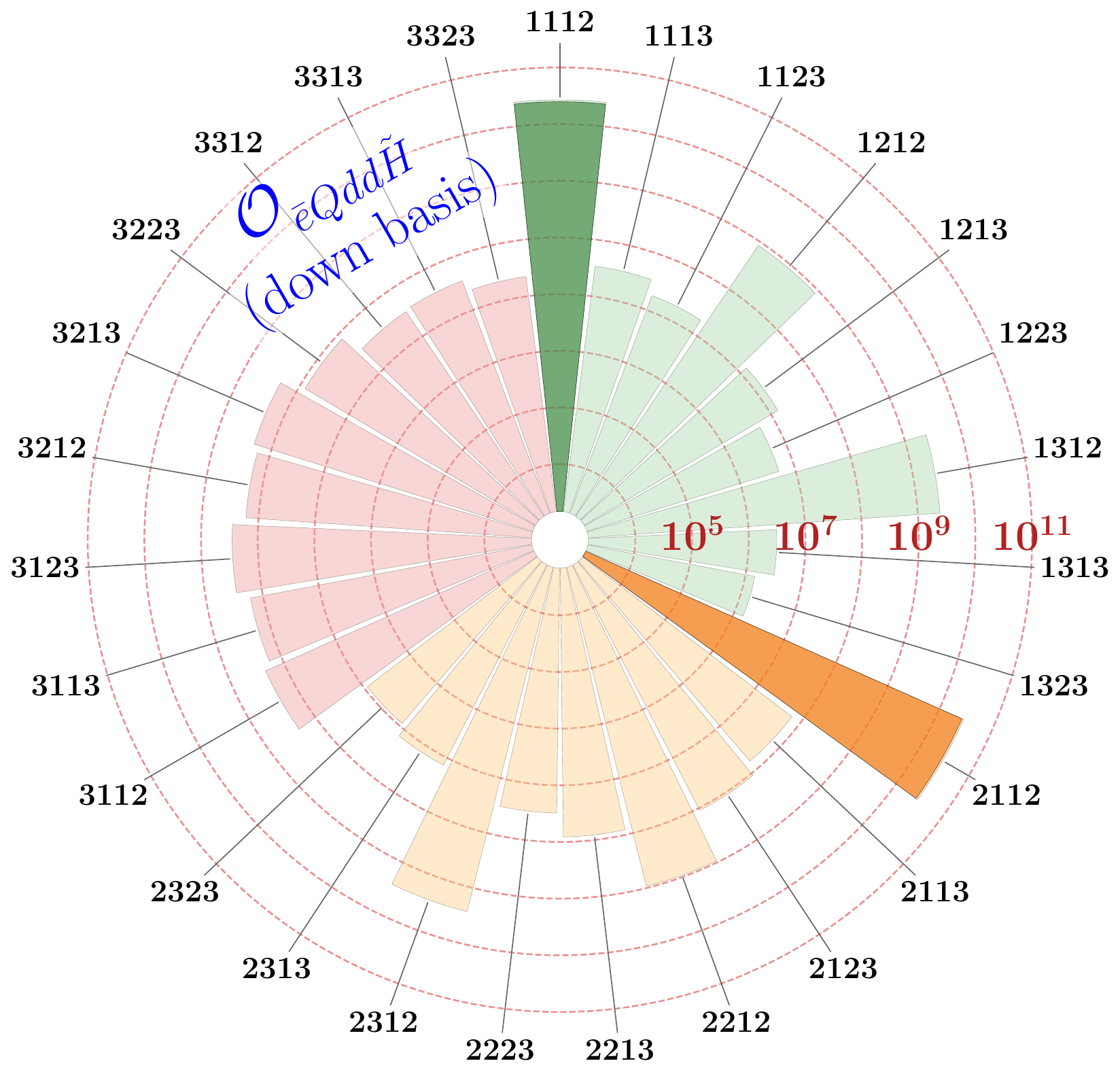}~
\includegraphics[width=0.31\linewidth]{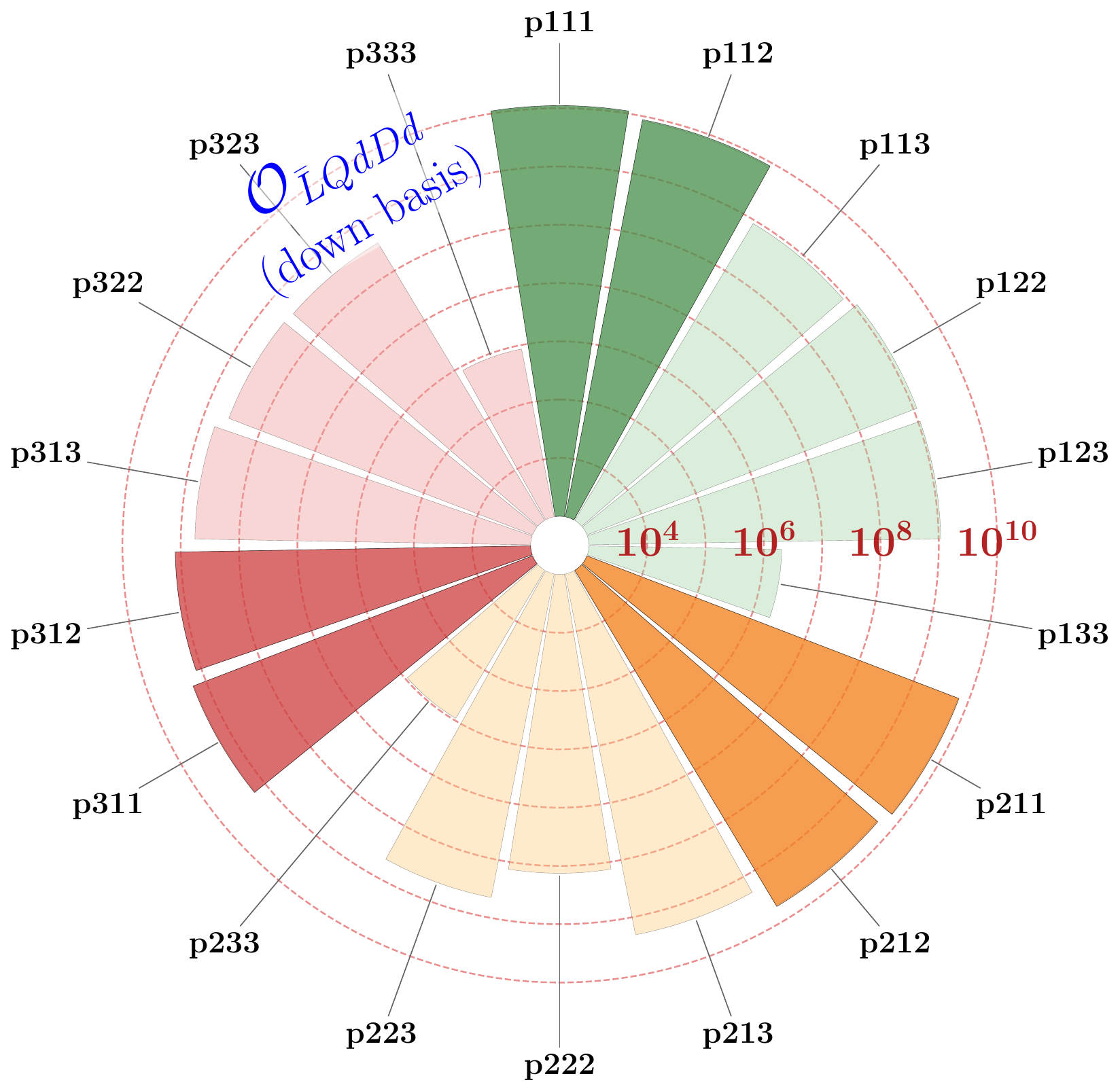}
\caption{Limits on the effective scales 
$\Lambda_{\tt eff}
=|C_7^i({\Lambda_{\tt NP})}|^{-1/3}$ 
of dim-7 SMEFT BNV operators in both the up-quark and down-quark flavor bases, derived from experimental bounds on two-body nucleon decays summarized in \cref{tab:exp_bound}. 
The light (dark) regions show the constraints with (without) SMEFT RG running effects, and different colors are used to distinguish between generation indices of a WC. 
For $\calO_{\bar{e}ddDd}^{p333}$, the black dashed and red solid arcs represent the results in the up- and down-quark flavor bases, respectively. 
Here, $p=1,~2,~3$ refers to the lepton generation. 
}
\label{fig:Bound_scale}
\end{figure*}

We present the results in two different ways: one in terms of an effective scale defined as $\Lambda_{\tt eff}
=|C_7^i ({\Lambda_{\tt NP})}|^{-1/3}$, and the other through a dimensionless WC given by $c_7^i\equiv |\Lambda_{\tt NP}^{3} C_7^i(\Lambda_{\tt NP})|$. 
The final bounds on $\Lambda_{\tt eff}$ are displayed in \cref{fig:Bound_scale}, while the constraints on $c_7^i$ are collected in \cref{tab:limit_dim7_Ops} for reference. 
Below, we provide a thorough examination of the results.

First, for the three operators $\calO_{\bar{L}dud\tilde H}$, $\calO_{\bar{L}dddH}$, and $\calO_{\bar{e}ddDd}$  (excluding the generation combinations $p333$), the constraints are practically the same in both quark-flavor bases. 
For the first two operators, this is because their dominant constraints arise from tree-level matching conditions and mixing structures, which are independent of the flavor basis choice to a good approximation. For $\calO_{\bar{e}ddDd}^{p333}$ with $p=1,~2,~3$, the results differ between the two bases because these operators primarily mix with $\calO_{\bar{L}dQQ\tilde H}$, whose contributions to the matching are basis-dependent.
In contrast, the constraints on the three operators $\calO_{\bar{L}dQQ\tilde H}$, $\calO_{\bar{e}Qdd\tilde H}$, and $\calO_{\bar{L}QdDd}$ exhibit flavor basis dependence. Because these operators involve the left-handed quark doublet,
their bounds are sensitive to the flavor basis choice, as is manifested in both matching conditions and RG evolution.

Second, the constraints on most of the generation combinations of the four operators involving the left-handed lepton doublet $L$ are independent of the lepton generation $p$ in \cref{fig:Bound_scale}.
This is because their most stringent bounds typically come from neutrino modes (see \cref{tab:exp_bound}), which do not distinguish between neutrino flavors.
The remaining operators discriminate the lepton flavor either by directly contributing to charged lepton modes involving the electron or muon, such as $\calO_{\bar{L}dddH}^{1112}$ and $\calO_{\bar{L}dddH}^{2112}$, or by mixing into operators associated with neutrino modes, like $\calO_{\bar{L}dddH}^{3122}$. 

Third, it is evident that the operators subject to the most stringent constraints are those that can directly match onto the LEFT nucleon decay operators, see the dark pie charts in \cref{fig:Bound_scale}.
This observation is consistent with our expectations. 
The effective scales associated with these operators are constrained in the range of $\calO(10^{9})-\calO(10^{11})$\,GeV, consistent with the previous tree-level analysis given in~\cite{Beneito:2023xbk,Gargalionis:2024nij}.
The RG-improved results for these operators are only slightly different from those without including RG evolution; for a clearer comparison, see the constraints on $c_7^i$ in \cref{tab:limit_dim7_Ops}.
For nearly all operators involving at least two second-generation quarks or one third-generation quark (the light pie charts in \cref{fig:Bound_scale}),  the constraints on $\Lambda_{\tt eff}$ vary from $\calO(10^5\,\rm GeV)$ to $\calO(10^{10}\,\rm GeV)$, depending on the specific generation combinations. 
These relatively weaker bounds are caused by their indirect contributions to the LEFT nucleon decay operators through RG mixings that are suppressed by small off-diagonal CKM matrix elements and Yukawa couplings.
Nevertheless, these indirect bounds remain significantly stronger than the direct limits from searches for BNV decays of the $\tau$ lepton~\cite{Heeck:2024jei}, hyperons~\cite{BESIII:2021slv}, and top quarks at LHC~\cite{Dong:2011rh,CMS:2024dzv}. 

In summary, our RG-improved analysis establishes stringent constraints on all independent dim-7 SMEFT BNV operators across various generation combinations,
representing a significant improvement over previous tree-level studies in the literature. 
Notably, for certain operators involving the second- and/or third-generation quarks, the RG-improved constraints become comparable in strength to those governing nucleon decays at tree level. 
For operators with relatively weaker constraints, further refinements may be achievable by incorporating the loop-induced matching contributions or by considering nucleon decay modes into multi-particle final states. Such processes could be assisted by the SM weak interactions that involve mediators like the $\tau$ lepton, the charm or bottom quark~\cite{Heeck:2024jei,Crivellin:2023ter,Beneke:2024hox,Chen:2025mjt}. 
On the other hand, current and upcoming neutrino experiments, including JUNO~\cite{JUNO:2015zny} Hyper-Kamiokande~\cite{Hyper-Kamiokande:2018ofw}, DUNE~\cite{DUNE:2020ypp}, and Theia~\cite{Theia:2019non}, 
are expected to further improve sensitivity to the nucleon decay processes in \cref{tab:exp_bound}. Their projected sensitivity would enhance the bounds on $c_7^i$ by about a factor of 2.

%%%%%%%%%%%%%%%%%%%%%%%%
\section{Summary}
\label{sec:summary}
%%%%%%%%%%%%%%%%%%%%%%%%

In this paper, we have presented an RG-improved analysis of the two-body $\Delta(B+L)=0$ nucleon decays that are induced by dim-7 SMEFT BNV interactions. By incorporating the complete one-loop RG equations, we were able to derive stringent constraints on all 297 independent Wilson coefficients across all possible generation combinations, extending beyond the previous tree-level analysis on operators involving at most one second-generation quark. 
For operators involving the $\tau$ lepton or top quark, the bounds on their effective scales exceed $\calO(10^5\,\rm GeV)$, significantly stronger than those obtained from the direct $\tau$ lepton and top quark decay searches. 
Our results underscore the crucial role of RG running effects in low-energy phenomenological studies, particularly the mixing effects driven by Yukawa interactions. 

%%%%%%%%%%%%%%%%%%%%%%%%
\acknowledgments
%%%%%%%%%%%%%%%%%%%%%%%%
This work was supported 
by Grants No.\,NSFC-12305110
and No.\,NSFC-12035008.  

\onecolumngrid
\appendix
\newlength{\fwidth}
\setlength{\fwidth}{0.3\textwidth}

%%%%%%%%%%%%%%%%%%%%%%%%
\section{Matching the derivative operator $\calO^{prst}_{\bar{L}QdDd}$ onto the LEFT}
\label{app:convert_LQddD}
%%%%%%%%%%%%%%%%%%%%%%%%

To match the operator $\mathcal{O}^{prst}_{\bar{L}QdDd}$ to the LEFT operators relevant to nucleon decays in~\cref{tab:matching_EW}, we employ integration by parts (IBP), the equations of motion (EoMs) of the SM fields, and Fierz identities.
Since 
\begin{align}
\calO^{prst}_{\bar{L}QdDd}=\frac{1}{2} 
\epsilon_{\alpha\beta\gamma}
\big(\overline{L_{p}} \gamma_\mu Q^{\alpha}_{r} \big)
\big(\overline{d^{\beta \C}_{s}} i D^\mu d^{\gamma}_{t} \big)+ s\leftrightarrow t, 
\end{align}
we only need to perform the manipulation for the first term on the right-hand side.
Using the following EoMs of the SM quark fields and Fierz identity~\cite{Liao:2016hru}:
\begin{subequations}
\begin{align}
i\slashed{D}d^{\alpha}_p&=(Y_{\rm d}^\dagger)_{pr}H^{\dagger}Q^{\alpha}_r,
\label{eq:EoMd}
\\
i\slashed{D}Q^{i\alpha}_p&=(Y_{\rm d})_{pr}H^i d_r^\alpha+(Y_{\rm u})_{pr}\tilde{H}^i u_r^\alpha,
\label{eq:EoMQ}
\\
(\overline{\Psi_{1\tL}}\gamma^\mu \Psi_{2\tL})(\overline{\Psi_{3\tR}^\C} \Psi_{4\tR})&=(\overline{\Psi_{1\tL}} \Psi_{3\tR})(\overline{\Psi_{2\tL}^\C}\gamma^\mu \Psi_{4\tR})+(\overline{\Psi_{1\tL}} \Psi_{4\tR})(\overline{\Psi_{2\tL}^\C}\gamma^\mu \Psi_{3\tR})\;,
\label{eq:FI}
\end{align}
\end{subequations}
we obtain
\begin{align}
\epsilon_{\alpha\beta\gamma}
\big(\overline{L_{p}} \gamma_\mu Q^{\alpha}_{r} \big)
\big(\overline{d^{\beta \C}_{s}} i D^\mu d^{\gamma}_{t} \big)  
 \overset{\eqref{eq:FI}}{=} &\epsilon_{\alpha\beta\gamma}
\big(\overline{L_{p}} d^{\beta }_{s}  \big)
\big(\overline{Q^{\alpha \C}_{r}} \gamma_\mu i D^\mu d^{\gamma}_{t} \big)
+\epsilon_{\alpha\beta\gamma}
\big(\overline{L_{p}} i D^\mu d^{\gamma}_{t}  \big)
\big(\overline{Q^{\alpha \C}_{r}} \gamma_\mu  d^{\beta }_{s}\big)
\nonumber\\
\overset{\eqref{eq:EoMd}}{=}&
-(Y_{\rm d}^{\dagger})_{tw}\epsilon_{ij}\epsilon_{\alpha\beta\gamma}
\big(\overline{L_{p}} d^{\beta }_{s}  \big)
\big(\overline{Q^{\alpha \C}_{r}} Q^{i\gamma}_{w} \big)\tilde{H}^j
+\epsilon_{\alpha\beta\gamma}
\big(\overline{L_{p}} i D^\mu d^{\gamma}_{t}  \big)
\big(\overline{Q^{\alpha \C}_{r}} \gamma_\mu  d^{\beta }_{s}\big)
\nonumber\\
\overset{\tt{IBP}}{=}~\,&
-(Y_{\rm d}^{\dagger})_{tw} \epsilon_{ij}\epsilon_{\alpha\beta\gamma}
\big(\overline{L_{p}} d^{\beta }_{s}  \big)
\big(\overline{Q^{\alpha \C}_{r}} Q^{i\gamma}_{w} \big)\tilde{H}^j
-\epsilon_{\alpha\beta\gamma}
\big(i \overline{D^\mu L_{p}}  d^{\gamma}_{t}  \big)
\big(\overline{Q^{\alpha \C}_{r}} \gamma_\mu  d^{\beta }_{s}\big)
\nonumber\\
&+\epsilon_{\alpha\beta\gamma}
\big(\overline{L_{p}}  d^{\gamma}_{t}  \big)
\big(\overline{d^{\beta \C }_{s}} \gamma_\mu i D^\mu Q^{\alpha }_{r} \big)-\epsilon_{\alpha\beta\gamma}
\big(\overline{L_{p}}  d^{\gamma}_{t}  \big)
\big(\overline{Q^{\alpha \C}_{r}} \gamma_\mu i D^\mu  d^{\beta }_{s}\big)
\nonumber\\
\overset{\eqref{eq:EoMQ}}{=}&-(Y_{\rm d}^{\dagger})_{tw}\epsilon_{ij}\epsilon_{\alpha\beta\gamma}
\big(\overline{L_{p}} d^{\beta }_{s}  \big)
\big(\overline{Q^{\alpha \C}_{r}} Q^{i\gamma}_{w} \big)\tilde{H}^j
-\epsilon_{\alpha\beta\gamma}
\big(i \overline{D^\mu L_{p}}  d^{\gamma}_{t}  \big)
\big(\overline{Q^{\alpha \C}_{r}} \gamma_\mu  d^{\beta }_{s}\big)
\nonumber\\
&+(Y_{\rm d})_{rw}\epsilon_{\alpha\beta\gamma}
\big(\overline{L_{p}}  d^{\gamma}_{t}  \big)
\big(\overline{d^{\beta \C }_{s}}  d^{\alpha }_{w} \big)H
+(Y_{\rm u})_{rw}\epsilon_{\alpha\beta\gamma}
\big(\overline{L_{p}}  d^{\gamma}_{t}  \big)
\big(\overline{d^{\beta \C }_{s}}  u^{\alpha }_{w} \big)\tilde{H}
\nonumber\\
&~~\,+(Y_{\rm d}^{\dagger})_{sw}\epsilon_{\alpha\beta\gamma}\epsilon_{ij}
\big(\overline{L_{p}}  d^{\gamma}_{t}  \big)
\big(\overline{Q^{\alpha \C}_{r}} Q^{\beta i}_{w}\big)\tilde{H}^j
\nonumber\\
=~~\,&(Y_{\rm d}^{\dagger})_{tw}\calO_{\bar{L}dQQ \tilde{H}}^{psrw}
+(Y_{\rm d}^{\dagger})_{sw}\calO_{\bar{L}dQQ \tilde{H}}^{ptrw}
-(Y_{\rm d})_{rw}\calO_{\bar{L}dddH}^{ptsw}
+(Y_{\rm u})_{rw}\calO_{\bar{L}dud\tilde{H}}^{ptws}
\nonumber\\
&-\epsilon_{\alpha\beta\gamma}
\big(i \overline{D^\mu L_{p}}  d^{\alpha}_{t}  \big)
\big(\overline{d^{\beta \C}_{s}} \gamma_\mu  Q^{\gamma }_{r}\big).
\label{eq:convert_LQddD}
\end{align}
From \cref{eq:convert_LQddD}, we finally obtain \cref{eq:OLQdDd_res},
\begin{align}
\calO^{prst}_{\bar{L} QdDd }=\, & 
\frac{1}{2}\big[
(Y_{\rm u})_{rw}\calO_{\bar{L}dud\tilde{H}}^{pswt}
-(Y_{\rm d})_{rw}\calO_{\bar{L}dddH}^{pstw}
+2 (Y_{\rm d}^{\dagger})_{sw}\calO_{\bar{L}dQQ \tilde{H}}^{ptrw}+ s\leftrightarrow t\big] 
-\epsilon_{\alpha\beta\gamma}
(i \overline{D^\mu L_{p}}  d^{\alpha}_{\{s})
(\overline{d^{\beta \C}_{t\}}} \gamma_\mu Q^{\gamma}_{r}).
\label{eq:OLQdDd_conversion}
\end{align}
Thus, its associated Lagrangian term becomes 
\begin{align}
C^{prst}_{\bar{L} QdDd } \calO^{prst}_{\bar{L} QdDd } =& 
\big[(Y_u)_{ws}C_{\bar{L}QdDd}^{pwtr}\big] 
\calO_{\bar{L}dud\tilde{H}}^{prst}
+\big[-(Y_d)_{wt}C_{\bar{L}QdDd}^{pwrs}\big]
\calO_{\bar{L}dddH}^{prst}
+\big[2(Y_d^\dagger)_{wt}C_{\bar{L}QdDd}^{psrw}\big] \calO_{\bar{L}dQQ\tilde{H}}^{prst}
\nonumber\\
& + \big[-C_{\bar{L}QdDd}^{ptrs} \big]
\big[\epsilon_{\alpha\beta\gamma}
(i \overline{D^\mu L_{p}}  d^{\alpha}_{\{r})
(\overline{d^{\beta \C}_{s\}}} \gamma_\mu Q^{\gamma}_{t})\big],
\label{eq:Lag_OLQdDd}
\end{align}
where we have used the flavor symmetry relation $C_{\bar{L}QdDd}^{prst}=C_{\bar{L}QdDd}^{prts}$, descending from the operator relation $\calO_{\bar{L}QdDd}^{prst}=\calO_{\bar{L}QdDd}^{prts}$.  
After taking into account the Higgs VEV and the CKM matrix elements, the three terms in the first line of \cref{eq:Lag_OLQdDd} modify the matching results for the three dim-7 operators that involve the Higgs field.
Meanwhile, the additional operator in the second line generates the matching results for the last two dim-7 operators with the lepton fields being acted upon by a derivative in~\cref{tab:matching_EW}.   

%%%%%%%%%%%%%%%%%%%%%%%%
\section{Chiral Lagrangian}
\label{app:chiral_matching}
%%%%%%%%%%%%%%%%%%%%%%%%

We organize in matrix form the ${\cal N}$s in the irreps ${\bf 8}_\tL\otimes {\bf 1}_\tR$ and $\bar{\pmb{3}}_\tL \otimes \pmb{3}_\tR$~\cite{Fan:2024gzc}:
\begin{subequations}
\label{eq:3qpart}
\begin{align}
{\cal N}_{\bar{\pmb{3}}_\tL \otimes \pmb{3}_\tR }  
=& \begin{pmatrix}
{\cal N}_{uds}^{\tR\tL}  
& {\cal N}_{usu}^{\tR\tL} 
& {\cal N}_{uud}^{\tR\tL} 
\\[1pt]
{\cal N}_{dds}^{\tR\tL}  
& {\cal N}_{dsu}^{\tR\tL} 
& {\cal N}_{dud}^{\tR\tL} 
\\[1pt] 
{\cal N}_{sds}^{\tR\tL}  
& {\cal N}_{ssu}^{\tR\tL} 
& {\cal N}_{sud}^{\tR\tL}
\end{pmatrix},\quad 
{\cal N}_{{\bf 8}_\tL\otimes {\bf 1}_\tR}
 =
\begin{pmatrix}
{\cal N}^{\tL\tL}_{uds}  &  {\cal N}^{\tL\tL}_{usu}  & {\cal N}^{\tL\tL}_{uud}  
\\[1pt]
{\cal N}^{\tL\tL}_{dds}  & {\cal N}^{\tL\tL}_{dsu} & {\cal N}^{\tL\tL}_{dud}  
\\[1pt]
{\cal N}^{\tL\tL}_{sds} & {\cal N}^{\tL\tL}_{ssu} & {\cal N}^{\tL\tL}_{sud}
\end{pmatrix}.
\end{align}
\end{subequations}
Their corresponding spurion matrices belong to the irreps ${\bf 8}_\tL\otimes {\bf 1}_\tR$ and $\pmb{3}_\tL \otimes \bar{\pmb{3}}_\tR$, and are denoted as follows: 
\begin{subequations}
\begin{align}
{\cal P}_{\pmb{3}_\tL \otimes \bar{\pmb{3}}_\tR }  
=& \begin{pmatrix}
{\cal P}_{uds}^{\tR\tL}  
&  {\cal P}_{dds}^{\tR\tL} 
& {\cal P}_{sds}^{\tR\tL} 
\\[2pt]
{\cal P}_{usu}^{\tR\tL}  
&  {\cal P}_{dsu}^{\tR\tL} 
& {\cal P}_{ssu}^{\tR\tL} 
\\[2pt] 
{\cal P}_{uud}^{\tR\tL} 
& {\cal P}_{dud}^{\tR\tL} 
& {\cal P}_{sud}^{\tR\tL}
\end{pmatrix}, \quad 
{\cal P}_{\pmb{8}_\tL \otimes \pmb{1}_\tR} = 
\begin{pmatrix}
0   & {\cal P}_{dds}^{\tL\tL}  
& {\cal P}_{sds}^{ \tL\tL} 
\\[2pt]
{\cal P}_{usu}^{\tL\tL}  
& {\cal P}_{dsu}^{\tL\tL} 
& {\cal P}_{ssu}^{\tL\tL} 
\\[2pt] 
{\cal P}_{uud}^{\tL\tL}  
& {\cal P}_{dud}^{\tL\tL} 
& {\cal P}_{sud}^{\tL\tL}
 \end{pmatrix}. 
\end{align}
\end{subequations}
Due to the trace condition ${\rm Tr}[{\cal N}_{{\bf 8}_\tL\otimes {\bf 1}_\tR}]=0$, 
we treat the operator ${\cal N}^{\tL\tL}_{uds}$ as redundant.
Accordingly, we set the corresponding (1,1) element of the $ {\cal P}_{\pmb{8}_\tL \otimes \pmb{1}_\tR}$ to zero, by incorporating its contribution into the (2,2) and (3,3) elements. Similar notations and conventions apply to their chirality partners with $\tL\leftrightarrow \tR$.

After taking the above spurion matrices into \cref{eq:LBlM} and expanding the pseudoscalar matrix to the zeroth order, we obtain the following relevant baryon-lepton mass mixing terms:
\begin{align}
{\cal L}_{Bl}^{{\tt \Delta( B+L)=0}} =& 
-\big(\hat{c}_1 C^{\tR \tL,x}_{\bar{\nu}dud} + \hat{c}_2 C^{\tR \tR,x}_{\bar{\nu}dud}\big)
(\overline{\nu_{\tL x}} n_{\tR} )
+\hat{c}_3\Lambda_\chi^{-1}C^{\tR \tL,x}_{\partial\bar{\nu}ddu}
(i\partial^\mu\overline{\nu_{\tL x}} i\tilde{\partial}_\mu n_{\tR} )
\nonumber\\
&- \frac{1}{\sqrt{6}}\big[\hat{c}_1 (C^{\tR \tL,x}_{\bar{\nu}dsu} - 2 C^{\tR \tL,x}_{\bar{\nu} sud})
+\hat{c}_2 (C^{\tR \tR,x}_{\bar{\nu}dsu}  - 2 C^{\tR \tR,x}_{\bar{\nu} sud}) \big]
(\overline{\nu_{\tL x}}\Lambda^0_{\tR} )
-\frac{3\hat{c}_3}{\sqrt{6}}\Lambda_\chi^{-1}C^{\tR \tL,x}_{\partial\bar{\nu}dsu}
( i\partial^\mu\overline{\nu_{\tL x}} i\tilde{\partial}_\mu \Lambda^0_{\tR} )
\nonumber\\
& + \frac{1}{\sqrt{2}} \big(\hat{c}_1 C^{\tR \tL,x}_{\bar{\nu}dsu} + 
\hat{c}_2 C^{\tR \tR,x}_{\bar{\nu}dsu} \big)
(\overline{\nu_{\tL x}} \Sigma^0_{\tR})
+\frac{\hat{c}_3}{\sqrt{2}}\Lambda_\chi^{-1}C^{\tR \tL,x}_{\partial\bar{\nu}dsu}
(i\partial^\mu\overline{\nu_{\tL x}} i\tilde{\partial}_\mu \Sigma^0_{\tR})
\nonumber\\
&+ \hat{c}_1 C^{\tL \tR,x}_{\bar{\ell} dds}
(\overline{\ell_{\tR x}}\Sigma^-_{\tL} )  
-\big(\hat{c}_1 C^{\tR \tL,x}_{\bar{\ell} dds}
+\hat{c}_2 C^{\tR \tR,x}_{\bar{\ell} dds} \big) 
(\overline{\ell_{\tL x}} \Sigma^-_{\tR}) 
-\hat{c}_3\Lambda_\chi^{-1}(C^{\tR \tL,x}_{\partial\bar{\ell}dds}-C^{\tR \tL,x}_{\partial\bar{\ell}dsd})
(i\partial^\mu\overline{\ell_{\tL x}} i\tilde{\partial}_\mu \Sigma^-_{\tR}),
\label{eq:LB2l}
\end{align}
where the QCD running effects in the LEFT have been incorporated into the LECs through $\hat c_{1,2} = 1.32 c_{1,2}$ and $\hat c_3= 0.91 c_3$, and
$\tilde{\partial}_\mu=\partial_\mu-\frac{1}{4}\gamma_\mu\slashed{\partial}$.
By expanding the chiral Lagrangian to the linear order in the meson fields, we obtain the relevant three-point interactions involving a nucleon, a meson, and a lepton, 
\begin{align}
{\cal L}_{\texttt{N}lM}^{{\tt \Delta( B+L)=0}} =\, &
 {\sqrt{2}i \over f_\pi} \Big\{ \frac{1}{\sqrt{2}} 
(\hat{c}_1 C^{\tR\tL,x}_{\bar\nu dud}+\hat{c}_2 C^{\tR\tR,x}_{\bar\nu dud}) 
(\overline{\nu_{\tL x}} p_{\tR})\pi^- 
-\frac{1}{\sqrt{2}}\frac{\hat{c}_3}{\Lambda_\chi}
C_{\partial\bar\nu ddu}^{\tR\tL,x}   (i\partial^\mu\overline{\nu_{\tL x}} i\tilde{\partial}_\mu p_{\tR})\pi^-
\nonumber
\\
&-{ 1 \over 2} (\hat{c}_1 C^{\tR\tL,x}_{\bar\nu dud} + \hat{c}_2 C^{\tR\tR,x}_{\bar\nu dud})
(\overline{\nu_{\tL x}} n_{\tR})\pi^0
+\frac{3}{2}\frac{\hat{c}_3}{\Lambda_\chi}C_{\partial\bar\nu ddu}^{\tR\tL,x}   (i\partial^\mu\overline{\nu_{\tL x}} i\tilde{\partial}_\mu n_{\tR})\pi^0 
\nonumber
\\
&- {1 \over 2\sqrt{3}} (\hat{c}_1 C^{\tR\tL,x}_{\bar\nu dud} - 3 \hat{c}_2 C^{\tR\tR,x}_{\bar\nu dud} ) (\overline{\nu_{\tL x}} n_{\tR}) \eta 
-\frac{1}{2\sqrt{3}}\frac{\hat{c}_3}{\Lambda_\chi}C_{\partial\bar\nu ddu}^{\tR\tL,x}   (i\partial^\mu\overline{\nu_{\tL x}} i\tilde{\partial}_\mu n_{\tR})\eta
\nonumber
\\
&+\frac{1}{\sqrt{2}} \big[\hat{c}_1 C^{\tL\tR,x}_{\bar\ell dds} (\overline{\ell_{\tR x}} n_{\tL})  +  (\hat{c}_1 C^{\tR\tL,x}_{\bar\ell dds} -\hat{c}_2 C^{\tR\tR,x}_{\bar\ell dds}) (\overline{\ell_{\tL x}} n_{\tR})\big] K^-
\nonumber
\\
&+\frac{1}{\sqrt{2}}\frac{\hat{c}_3}{\Lambda_\chi}(C_{\partial\bar\ell dds}^{\tR\tL,x}+C_{\partial\bar\ell dsd}^{\tR\tL,x})   (i\partial^\mu\overline{\ell_{\tL x}} i\tilde{\partial}_\mu n_{\tR})K^-
\nonumber
\\
&+ \frac{1}{\sqrt{2}} ( \hat{c}_1 C^{\tR\tL,x}_{\bar\nu sud} + \hat{c}_2 C^{\tR\tR,x}_{\bar\nu sud} ) (\overline{\nu_{\tL x}} p_{\tR}) K^- 
-\frac{1}{\sqrt{2}}\frac{\hat{c}_3}{\Lambda_\chi}C_{\partial\bar\nu dsu}^{\tR\tL,x}   (i\partial^\mu\overline{\nu_{\tL x}} i\tilde{\partial}_\mu p_{\tR})K^- 
\nonumber
\\
&+\frac{1}{\sqrt{2}} \left[ \hat{c}_1( C^{\tR\tL,x}_{\bar\nu sud}  + C^{\tR\tL,x}_{\bar\nu dsu}) + \hat{c}_2 (C^{\tR\tR,x}_{\bar\nu sud} - C^{\tR\tR,x}_{\bar\nu dsu}) \right] (\overline{\nu_{\tL x}} n_{\tR}) \bar{K}^0
\nonumber
\\
&-\frac{\sqrt{2}\hat{c}_3}{\Lambda_\chi}C_{\partial\bar\nu dsu}^{\tR\tL,x}   (i\partial^\mu\overline{\nu_{\tL x}} i\tilde{\partial}_\mu n_{\tR})\bar{K^0} 
+\frac{\sqrt{2}\hat{c}_3}{\Lambda_\chi}C_{\partial\bar\ell ddd}^{\tR\tL,x}   (i\partial^\mu\overline{\ell_{\tL x}} i\tilde{\partial}_\mu n_{\tR})\pi^-
\Big\}.
\label{eq:NlM_contact}
\end{align}

In addition to the BNV interactions given above, the standard leading-order chiral interactions involving baryons are also required for the non-contact contributions to $\texttt{N}\to l M$ shown as a black square in \cref{fig:Feyndiagram}. These interactions have been worked out previously and we summarize the relevant terms as follows~\cite{Jenkins:1990jv,Bijnens:1985kj,Ma:2025mjy},
\begin{align}
{\cal L}_{\bar{B}\texttt{N}M} & \supset 
\frac{D-F}{\sqrt{2}f_\pi}
 \left[ 
  \overline{\Sigma^0} \gamma^\mu \gamma_5 p \, \partial_\mu K^-
- \overline{\Sigma^0} \gamma^\mu \gamma_5 n\, \partial_\mu \bar K^0 
+ \sqrt{2}\big(\overline{\Sigma^+}\gamma^\mu \gamma_5 p \, \partial_\mu \bar K^0 
+ \overline{\Sigma^-} \gamma^\mu \gamma_5 n \, \partial_\mu K^- \big) 
\right]
\notag  \\ 
& 
+ \frac{3F-D}{\sqrt{6}f_\pi} 
\big(
  \overline{p} \gamma^\mu \gamma_5 p \, \partial_\mu \eta 
+ \overline{n} \gamma^\mu \gamma_5 n \, \partial_\mu \eta \big)
-\frac{D+3F}{\sqrt{6}f_\pi} 
\left[ 
  \overline{\Lambda^0} \gamma^\mu \gamma_5 p \, \partial_\mu K^- 
+ \overline{\Lambda^0} \gamma^\mu \gamma_5 n \, \partial_\mu \bar K^0
\right]
\notag  \\ 
&
+\frac{D+F}{\sqrt{2}f_\pi} 
\left[
  \overline{p} \gamma^\mu \gamma_5 p\, \partial_\mu \pi^0
- \overline{n} \gamma^\mu \gamma_5 n\, \partial_\mu \pi^0
+ \sqrt{2} \big(\overline{n} \gamma^\mu \gamma_5 p \, \partial_\mu \pi^- 
+ \overline{p} \gamma^\mu \gamma_5 n \, \partial_\mu \pi^+ \big)
\right].
\label{eq:LBBM}
\end{align}
We use the low-energy constants $D=0.730(11)$ and $F=0.447^{6}_{7}$ from the recent lattice calculation~\cite{Bali:2022qja}.

\begin{table}[t]
\center
\resizebox{\linewidth}{!}{
\renewcommand{\arraystretch}{2.1}
\begin{tabular}{|c| c| c | c | c |c |}
\hline
$\texttt{N}\to lM$
& $C_{\texttt{N} \to BM}$
& \multicolumn{1}{c|}{$C^{1\tL(\tR)}_{Bl}$}
& \multicolumn{1}{c|}{$C^{1\tL(\tR)}_{\texttt{N}lM}$} 
& \multicolumn{1}{c|}{$C^{3\tR}_{Bl}$}
& \multicolumn{1}{c|}{$C^{3\tR}_{\texttt{N}lM}$} 
\\\hline
$p \to \nu_x \pi^+ $
& $\makecell{C_{p\to n\pi^+}\\=D+F }$
& $C^{1\tR}_{n\nu}=-(\hat c_1 C^{\tR\tL,x}_{\bar{\nu}dud}
+ \hat c_2 C^{\tR \tR,x}_{\bar{\nu}dud})$
& $C^{1\tR}_{p\nu \pi^+}= 
\hat c_1 C_{\bar\nu dud}^{\tR\tL,x} 
+ \hat c_2 C_{\bar\nu dud}^{\tR\tR,x}$
&$C^{3\tR}_{n\nu}=\hat c_3 
C_{\partial\bar\nu ddu}^{\tR\tL,x}$
& $C^{3\tR}_{p\nu \pi^+}=-\hat c_3 
C_{\partial\bar\nu ddu}^{\tR\tL,x} $
\\\hline 
\multirow{2}*{$ p \to \nu_x K^+$}
&$\makecell{C_{p\to \Lambda^0 K^+} \\[2pt]
=-\frac{D+3F}{\sqrt{6}} }$
& \makecell{$C^{1\tR}_{\Lambda^0 \nu}=-\frac{1}{\sqrt{6}} \big[\hat c_1 (C^{\tR \tL,x}_{\bar{\nu}dsu} 
- 2 C^{\tR \tL,x}_{\bar{\nu} sud} ) $\\[4pt]
$+ \hat c_2 (C^{\tR \tR,x}_{\bar{\nu}dsu} 
- 2 C^{\tR \tR,x}_{\bar{\nu} sud} ) \big]$}
& \multirow{2}*{$\makecell{C^{1\tR}_{p\nu K^+}=
\hat c_1 C^{\tR\tL,x}_{\bar\nu sud} 
+ \hat c_2 C^{\tR\tR,x}_{\bar\nu sud}}$}
&{$C^{3\tR}_{\Lambda^0\nu}=- \frac{3}{\sqrt{6}}
\hat c_3 C_{\partial\bar\nu dsu}^{\tR\tL,x} $}
& \multirow{2}*{$C^{3\tR}_{p\nu K^+}=- \hat c_3 C_{\partial\bar\nu dsu}^{\tR\tL,x} $}
\\\hhline{~--~-}
& $\makecell{C_{p\to \Sigma^0 K^+}\\
=\frac{D-F}{\sqrt{2} } }$
& $C^{1\tR}_{\Sigma^0 \nu}=\frac{1}{\sqrt{2}} 
(\hat c_1 C^{\tR \tL,x}_{\bar{\nu}dsu} 
+ \hat c_2 C^{\tR \tR,x}_{\bar{\nu}dsu})$
& 
& $C^{3\tR}_{\Sigma^0 \nu}=\frac{1}{\sqrt{2}}
\hat c_3 C_{\partial\bar\nu dsu}^{\tR\tL,x} $
&
\\\hline %n->nu pi0
{$n \to \nu_x \pi^0 $}
& $\makecell{C_{n\to n \pi^0}\\=-\frac{D+F}{\sqrt{2}} }$
& $C_{n\nu}^{1\tR}$
& {$C^{1\tR}_{n\nu \pi^0}=-{ 1 \over \sqrt{2}} 
(\hat c_1 C^{\tR\tL,x}_{\bar\nu dud} 
+ \hat c_2 C^{\tR\tR,x}_{\bar\nu dud})$}
&$C^{3\tR}_{n\nu}$
&{$C^{3\tR}_{n\nu \pi^0}={3\over \sqrt{2}}
\hat c_3 C_{\partial\bar\nu ddu}^{\tR\tL,x} $}
\\\hline %n->nu eta
{$n \to \nu_x \eta $}
& $\makecell{C_{n\to n \eta}\\
=-\frac{D-3F}{\sqrt{6}} }$
& $C_{n\nu}^{1\tR}$
&~{$C^{1\tR}_{n\nu\eta}=- \frac{1}{\sqrt{6}} 
(\hat c_1 C^{\tR\tL,x}_{\bar\nu dud} 
- 3 \hat c_2 C^{\tR\tR,x}_{\bar\nu dud} )$}~
&$C^{3\tR}_{n\nu}$
&{$C^{3\tR}_{n\nu\eta}=- \frac{1}{\sqrt{6}} 
\hat c_3 C_{\partial\bar\nu ddu}^{\tR\tL,x} $}
\\\hline %n->nu K0
\multirow{2}*{$n \to \nu_x K^0$}
& $\makecell{C_{n\to \Lambda^0 K^0}\\
=-\frac{D+3F}{\sqrt{6}} }$
& $C_{\Lambda^0\nu}^{1\tR}$
& \multirow{2}*{$ \makecell{C^{1\tR}_{n\nu K^0}=
\hat c_1( C^{\tR\tL,x}_{\bar\nu sud}  
+ C^{\tR\tL,x}_{\bar\nu dsu}) 
\\[4pt]
\quad\quad\quad\quad+ \hat c_2 (C^{\tR\tR,x}_{\bar\nu sud} 
- C^{\tR\tR,x}_{\bar\nu dsu}) }$ }
& $C^{3\tR}_{\Lambda^0\nu}$
&\multirow{2}*{$C^{3\tR}_{n\nu K^0}=
-2 \hat c_3 C_{\partial\bar\nu dsu}^{\tR\tL,x} $}
\\\hhline{~--~-~}
& $\makecell{C_{n\to \Sigma^0 K^0}\\
=-\frac{D-F}{\sqrt{2} } }$
& $C_{\Sigma^0\nu}^{1\tR}$
& 
& $C^{3\tR}_{\Sigma^0\nu}$
&
\\\hline %n->ell- pi+
$n\to \ell_x^- \pi^+$ 
& --
& --
& --
&--
&$C^{3\tR}_{n\ell^-\pi^+}=
2 \hat c_3  C_{\partial\bar\ell ddd}^{\tR\tL,x}$
\\\hline %n->ell- K+
~$n\to \ell_x^- K^+$~
& $\makecell{~C_{n\to \Sigma^- K^+}\\=D-F}$
& $\makecell{
C^{1\tL}_{\Sigma^-\ell^-}=\hat c_1 C^{\tL \tR,x}_{\bar{\ell} dds} \\[4pt]
C^{1\tR}_{\Sigma^-\ell^-}=
-(\hat c_1 C^{\tR \tL,x}_{\bar{\ell}dds}
+ \hat c_2 C^{\tR \tR,x}_{\bar{\ell}dds})}$
&$\makecell{ C^{1\tL}_{n\ell^-K^+}=
\hat c_1 C^{\tL\tR,x}_{\bar\ell dds} \\[4pt]
C^{1\tR}_{n\ell^-K^+}=
\hat c_1 C^{\tR\tL,x}_{\bar\ell dds} 
-\hat c_2 C^{\tR\tR,x}_{\bar\ell dds} } $
&~$C^{3\tR}_{\Sigma^-\ell^-}= 
\hat c_3 (C_{\partial\bar\ell dsd}^{\tR\tL,x} -C_{\partial\bar\ell dds}^{\tR\tL,x}) $~
&~$C^{3\tR}_{n\ell^-K^+}=\hat c_3 (C_{\partial\bar\ell dds}^{\tR\tL,x}+C_{\partial\bar\ell dsd}^{\tR\tL,x}) $~
\\\hline
\end{tabular} }
\caption{Summary of the coefficients in \cref{eq:LB2lM}.}
\label{tab:B2lM_vertex}
\end{table}

In general, the interactions contributing to each process ${\tt N}\to l M$ can be parametrized as follows, 
\begin{align}
 {\cal L}_{\texttt{N}\to l M}
=\, &
 \frac{ C_{\texttt{N}\to BM}}{f_\pi} 
 \overline{B}\gamma_\mu \gamma_5 \texttt{N} \, \partial^\mu \bar M +
 C_{Bl}^{1\tL} \bar{L}B_\tL+ C_{Bl}^{1\tR}\bar{L}B_\tR
+ \frac{C_{Bl}^{3\tL}}{\Lambda_\chi} i\partial^\mu\bar{L}i\tilde{\partial}_\mu B_\tL
+ \frac{C_{Bl}^{3\tR}}{\Lambda_\chi} i\partial^\mu\bar{L}i\tilde{\partial}_\mu B_\tR
\nonumber\\
& + \frac{i}{f_\pi} \Big(C_{{\tt N}lM}^{1\tL} \bar M \bar{L} {\tt N}_\tL
+ C_{{\tt N}lM}^{1\tR} \bar M \bar{L} {\tt N}_\tR
+ \frac{C_{{\tt N}lM}^{3\tR}}{\Lambda_\chi} \bar Mi\partial^\mu\bar{L}i\tilde{\partial}_\mu{\tt N}_\tR
+ \frac{C_{{\tt N}lM}^{3\tL}}{\Lambda_\chi} \bar Mi\partial^\mu\bar{L}i\tilde{\partial}_\mu{\tt N}_\tL
\Big),
\label{eq:LB2lM}
\end{align}
where the terms in the first line correspond to the non-contact contributions shown in \cref{fig:Feyndiagram}.
For a given field configuration, the coefficients $C_{\texttt{N}\to BM}$s and $C_{Bl}^{1(3)\tL(\tR)}$s can be obtained from \cref{eq:LBBM} and \cref{eq:LB2l}, respectively. 
The terms in the second line contribute to the contact diagram in \cref{fig:Feyndiagram}, with the coefficients
$C_{{\tt N}lM}^{(1)3\tL(\tR)}$ easily read off from \cref{eq:NlM_contact}.
All relevant coefficients are summarized in \cref{tab:B2lM_vertex} for each process; coefficients that were not matched from the dim-7 SMEFT BNV interactions are omitted. 
By evaluating the two diagrams shown in \cref{fig:Feyndiagram}, we obtain the following general expression for the decay width,
\begin{align}
\label{eq:GammaN2lM}
\Gamma_{{\tt N} \to l M} = \frac{m_{\tt N} \lambda^{1/2}(1,x_l, x_M)} {32\pi f_\pi^2 }
\Big[ (1 + x_l -x_M)
\big( |\tilde C_{{\tt N}lM}^\tL|^2 
+ |\tilde C_{{\tt N}lM}^\tR|^2\big) 
+ 4 x_l^{1\over2}\Re(\tilde C_{{\tt N}lM}^\tL \tilde C_{{\tt N}lM}^{\tR,*}) \Big],
\end{align}
where $\lambda(x,y,z)\equiv x^2+y^2+z^2-2xy-2yz-2zx$ is the triangle function.
$x_l=m_l^2/m_{\tt N}^2$, $x_M=m_M^2/m_{\tt N}^2$, and
$\tilde C_{{\tt N}lM}^{\tL,\tR}$ are related to the parameters in \cref{eq:LB2lM} via
\begin{subequations}
\begin{align}
\tilde C_{{\tt N}lM}^\tL  
=\,& C_{{\tt N}lM}^{1\tL}  
+\frac{C_{{\tt N}lM}^{3\tR}}{4\Lambda_\chi} m_\N m_l 
-\frac{C_{{\tt N}lM}^{3\tL}}{2\Lambda_\chi}(m_\N^2+m_l^2-m_M^2)
+\sum_B \frac{C_{{\tt N}\to BM}}{m_B^2 - m_l^2} 
\big[ (m_{\tt N} m_B + m_l^2) C_{Bl}^{1\tL} + m_l(m_{\tt N} +m_B) C_{Bl}^{1\tR} 
\nonumber\\
& -\frac{3}{4}\frac{C_{Bl}^{3\tL}}{\Lambda_\chi}m_l^2(m_B m_\N+m_l^2)
-
\frac{3}{4}\frac{C_{Bl}^{3\tR}}{\Lambda_\chi}m_l^3(m_B+ m_\N)\big],
\\
\tilde C_{{\tt N}lM}^\tR  
=\,& C_{{\tt N}lM}^{1\tR}
+\frac{C_{{\tt N}lM}^{3\tL}}{4\Lambda_\chi} m_\N m_l
-\frac{C_{{\tt N}lM}^{3\tR}}{2\Lambda_\chi}(m_\N^2+m_l^2-m_M^2)
- \sum_B \frac{ C_{{\tt N}\to BM} }{m_B^2 - m_l^2} 
\Big[ (m_{\tt N} m_B + m_l^2) C_{Bl}^{1\tR} + m_l(m_{\tt N} +m_B) C_{Bl}^{1\tL}
\nonumber\\
&-\frac{3}{4}\frac{C_{Bl}^{3\tR}}{\Lambda_\chi}m_l^2(m_B m_\N+m_l^2)
-\frac{3}{4}\frac{C_{Bl}^{3\tL}}{\Lambda_\chi}m_l^3(m_B+ m_\N)\Big].
\end{align}
\label{eq:Dewidth_express_LEFT}
\end{subequations}

%%%%%%%%%%%%%%%%%%%%%%%%
\section{Constraints on the dimensionless Wilson coefficients}
\label{app:bound_on_c7i}
%%%%%%%%%%%%%%%%%%%%%%%%

\begin{table}[H]
\center
\resizebox{0.9\textwidth}{!}{
\renewcommand{\arraystretch}{1.05}
\begin{tabular}{|c|c|c|c|c|c|c|c|c|}
\hline 
\multicolumn{9}{|c|}{
\large RGE-improved constraints on the dimensionless WCs $c_7^i\equiv |\Lambda_{\tt NP}^{3} C_7^i(\Lambda_{\tt NP})|$ at $\Lambda_{\tt NP}=10^8$ GeV}
\\\hline
~~\makecell{Generation \\ indices}~~
&~~Up-basis~~
&~~Down-basis~~
&~~\makecell{Generation \\ indices}~~
&~~Up-basis~~
&~~Down-basis~~
&~~\makecell{Generation \\ indices}~~
&~~Up-basis~~
&~~Down-basis~~
\\\cline{1-9}
\multicolumn{3}{|c|}{\large$\calO_{\bar{e}Qdd\tilde{H}}$}
& \multicolumn{3}{c|}{\large$\calO_{\bar{L}QdDd}$}
& \multicolumn{3}{c|}{\large$\calO_{\bar{L}dud\tilde{H}}$}
\\\cline{1-9}
$1112$ &$5.3(6.7)\cdot10^{-9}$ &$5.2(6.5)\cdot10^{-9}$
& $p111$ &$7.9(7.1)\cdot10^{-7}$ &$8.1(7.3)\cdot10^{-7}$ 
&$p111$  &\multicolumn{2}{c|}{$8.5(11)\cdot10^{-10}$} 
\\
$1113$ &$2.6$ &$2.5$
& $p112$ &$1.7(1.5)\cdot10^{-6}$ &$1.7(1.5)\cdot10^{-6}$ 
&$p112$  &\multicolumn{2}{c|}{$1.7(2.6)\cdot10^{-9}$}
\\
$1123$ &$15.4$ &$15.5$  
& $p113$ 
&\multicolumn{2}{c|}{
$1.5\cdot10^{-3}$}
&$p113$  &\multicolumn{2}{c|}{$0.85$}
\\
$1212$ &$2.3(2.9)\cdot10^{-8}$ &$1.6\cdot10^{-4}$
& $p122$ &$6.7\cdot10^{-4}$ &$6.5\cdot10^{-4}$ 
&$p121$  &\multicolumn{2}{c|}{$0.22$}
\\
$1213$ &$11$ &$38$
& $p123$ &$8.0\cdot10^{-4}$ &$7.9\cdot10^{-4}$ 
&$p122$  &\multicolumn{2}{c|}{$0.078$} 
\\
$1223$ &$94$ &$710$
& $p133$ &$4.2\cdot10^{3}$ &$1.2\cdot10^{5}$ 
&$p123$  &\multicolumn{2}{c|}{$0.028$} 
\\
$1312$ &$5.2(7.0)\cdot10^{-7}$ &$6.8\cdot10^{-6}$
& $1211$ &$1.7(1.5)\cdot10^{-5}$ &$3.5(3.2)\cdot10^{-6}$ 
&$p131$  &\multicolumn{2}{c|}{$0.020$} 
\\
$1313$ &$240$ &$3.3\cdot10^{3}$
& $2211$ &$1.3(1.1)\cdot10^{-5}$ &$3.5(3.2)\cdot10^{-6}$ 
&$p132$  &\multicolumn{2}{c|}{$7.0\cdot10^{-3}$}
\\
$1323$ &$2.3\cdot10^{3}$ &$3.2\cdot10^{4}$
& $3211$ &$0.012$ &$3.5(3.2)\cdot10^{-6}$ 
&$p133$  &\multicolumn{2}{c|}{$2.4\cdot10^{-3}$}
\\
$2112$ &$4.1(5.1)\cdot10^{-9}$ &$4.0(5.0)\cdot10^{-9}$
& $1212$ &$5.3(4.3)\cdot10^{-5}$ &$7.6(6.5)\cdot10^{-6}$ 
&$p211$  &\multicolumn{2}{c|}{$5.4(7.1)\cdot10^{-10}$}
\\
$2113$ &$0.37$ &$0.36$
& $2212$ &$4.0(3.3)\cdot10^{-5}$ &$7.6(6.5)\cdot10^{-6}$ 
&$p212$  &\multicolumn{2}{c|}{$6.3$}
\\
$2123$ &\multicolumn{2}{c|}{$0.075$}
& $3212$ &$3.0\cdot10^{-3}$ &$7.6(6.5)\cdot10^{-6}$ 
&$p213$  &\multicolumn{2}{c|}{$11.4$} 
\\
$2212$ &$1.8(2.2)\cdot10^{-8}$ &$1.2\cdot10^{-4}$
& $p213$ &$1.0\cdot10^{-3}$ &$1.2\cdot10^{-4}$ 
&$p221$  &\multicolumn{2}{c|}{$0.18$}  
\\
$2213$ &$0.14$ &$0.19$
& $p222$ &$2.9\cdot10^{-3}$ &$0.42$ 
&$p222$  &\multicolumn{2}{c|}{$0.055$}
\\
$2223$ &$0.45$ &$3.5$
& $p223$ &$3.3\cdot10^{-3}$ &$0.011$ 
&$p223$  &\multicolumn{2}{c|}{$0.093$}
\\
$2312$ &$3.9(5.3)\cdot10^{-7}$ &$5.2\cdot10^{-6}$
& $p233$ &$1.8\cdot10^{4}$ &$1.3\cdot10^{6}$ 
&$p231$  &\multicolumn{2}{c|}{$0.017$}
\\
$2313$ &$2.7$ &$38$
& $1311$ &$4.2(3.6)\cdot10^{-4}$ &$2.2(1.9)\cdot10^{-4}$ 
&$p232$  &\multicolumn{2}{c|}{$5.3\cdot10^{-3}$}
\\
$2323$ &$11$ &$150$
& $2311$ &$3.2(2.7)\cdot10^{-4}$ &$2.2(1.9)\cdot10^{-4}$ 
&$p233$  &\multicolumn{2}{c|}{$9.8\cdot10^{-3}$}
\\
$3112$ &$9.1\cdot10^{-3}$ &$8.8\cdot10^{-3}$
& $3311$ &$0.054$ &$2.2(1.9)\cdot10^{-4}$ 
&$p311$  &\multicolumn{2}{c|}{$0.26$}
\\
$3113$ &$0.022$ &$0.021$
& $1312$ &$1.4(1.1)\cdot10^{-3}$ &$5.3(5.2)\cdot10^{-4}$ 
&$p312$  &\multicolumn{2}{c|}{$19.3$}
\\
$3123$ &$4.4\cdot10^{-3}$ &$4.5\cdot10^{-3}$
& $2312$ &$10(8.3)\cdot10^{-4}$ &$5.3(5.2)\cdot10^{-4}$ 
&$p313$  &\multicolumn{2}{c|}{$4.7\cdot10^9$}
\\
$3212$ &$0.017$ &$0.022$
& $3312$ &$0.014$ &$5.3(5.2)\cdot10^{-4}$ 
&$p321$  &\multicolumn{2}{c|}{$0.055$}
\\
$3213$ &$8.3\cdot10^{-3}$ &$0.011$
& $p313$ &$4.4\cdot10^{-3}$ &$5.3\cdot10^{-3}$ 
&$p322$  &\multicolumn{2}{c|}{$0.18$}
\\
$3223$ &$0.027$ &$0.21$
& $p322$ &$0.015$ &$0.018$ 
&$p323$  &\multicolumn{2}{c|}{$4.3\cdot10^7$}
\\
$3312$ &$0.37$ &$2.7$
& $p323$ &$0.018$ &$0.021$ 
&$p331$  &\multicolumn{2}{c|}{$4.7\cdot10^{-3}$} 
\\
$3313$ &$0.16$ &$2.2$
& $p333$ &$6.9\cdot10^{4}$ &$1.6\cdot10^{6}$ 
&$p332$  &\multicolumn{2}{c|}{$0.019$}
\\
$3323$ &$0.65$ &$9.1$
&  & &
&$p333$  &\multicolumn{2}{c|}{$5.1\cdot10^6$}
\\\cline{1-9}
%%%%%%%%%%%%%%%%%%%%%%%%%%%%%%%%%%%%%%%%%%%%%
\multicolumn{6}{|c|}{\large$\calO_{\bar{L}dQQ\tilde{H}}$}
&\multicolumn{3}{c|}{\large$\calO_{\bar{L}dddH}$}
\\\cline{1-9}
$p111$
& \multicolumn{2}{c|}{$7.9(12) \cdot 10^{-10}$} 
& $p212$ 
& $2.3(3.2)\cdot10^{-9}$
& $2.3\cdot10^{-8}$
&$1112$
&\multicolumn{2}{c|}{$8.7(11)\cdot10^{-9}$}
\\
$p112$
& \multicolumn{2}{c|}{$2.0(2.7) \cdot 10^{-9}$}
& $p213$ 
& $5.4(7.9)\cdot10^{-8}$
& $7.1\cdot10^{-7}$
&$2112$
&\multicolumn{2}{c|}{$6.4(8.3)\cdot10^{-9}$}
\\
$p113$
& $4.3(6.3) \cdot 10^{-8}$ 
& $6.2 \cdot 10^{-7}$
& $p221$ 
& $2.3\cdot10^{-8}$
& $2.3(3.2)\cdot10^{-9}$
&$3112$
&\multicolumn{2}{c|}{$16$}
\\
$1121$
& $1.1(1.3)\cdot10^{-8}$ 
& $3.8(5.1) \cdot 10^{-9}$
& $p222$ 
& $1.4$
& $2.2\cdot10^{-4}$
&$p212$
&\multicolumn{2}{c|}{$5.8$}
\\
$2121$
& $8.1(9.9)\cdot10^{-9}$ 
& $3.8(5.1) \cdot 10^{-9}$
& $p223$ 
& $0.32$
& $3.4\cdot10^{-6}$
&$p312$
&\multicolumn{2}{c|}{$4.7$}
\\
$3121$
& $1.9\cdot10^{-8}$ 
& $3.8(5.1) \cdot 10^{-9}$
& $p231$ 
& $2.1\cdot10^{-5}$
& $4.0\cdot10^{-8}(1.8\cdot10^{-7})$
&$1113$
&\multicolumn{2}{c|}{$5.2$}
\\
$p122$
& $0.95$ 
& $7.8(12) \cdot 10^{-9}$
& $p232$ 
& $0.13$
& $2.5\cdot10^{-6}$
&$2113$
&\multicolumn{2}{c|}{$3.9$}
\\
$1123$
& $1.3(1.8) \cdot 10^{-6}$
& $2.8 \cdot 10^{-6}$
& $p233$ 
& $5.7$
& $5.9\cdot10^{-4}$
&$3113$
&\multicolumn{2}{c|}{$21$}
\\
$2123$
& $9.8(13) \cdot 10^{-7}$ 
& $2.8 \cdot 10^{-6}$
& $p311$ 
& \multicolumn{2}{c|}{$0.24$}
&$p213$
&\multicolumn{2}{c|}{$6.4$}
\\
$3123$
& $0.097$ 
& $2.8 \cdot 10^{-6}$
& $p312$ 
& \multicolumn{2}{c|}{$1.1$}
&$p313$
&\multicolumn{2}{c|}{$150$} 
\\
$1131$
& $2.6(3.1) \cdot 10^{-7}$ 
& $1.7(5.1) \cdot 10^{-7}$
& $p313$ 
& $26$ & $340$
&$p223$
&\multicolumn{2}{c|}{$6.5$}
\\
$2131$
& $2.0(2.3) \cdot 10^{-7}$ 
& $1.7(5.1) \cdot 10^{-7}$
& $p321$ 
& $0.44$ & $0.46$
&$p323$
&\multicolumn{2}{c|}{$620$}
\\
$3131$
& $1.7 \cdot 10^{-5}$ 
& $1.7(5.1) \cdot 10^{-7}$
& $p322$ 
& $2.1$ & $21$
&
\multicolumn{3}{c|}{}
\\
$1132$
& $1.5(1.8) \cdot10^{-6}$ 
& $2.9(2.4) \cdot 10^{-7}$
& $p323$ 
& $50$ & $620$
&
\multicolumn{3}{c|}{}
\\
$2132$
& $1.1(1.3)\cdot10^{-6}$ 
& $2.9(2.4) \cdot 10^{-7}$
& $p331$ 
& $9.2$ & $73$
&
\multicolumn{3}{c|}{}
\\
$3132$
& $0.039$ 
& $2.9(2.4) \cdot 10^{-7}$
& $p332$ 
& $57$ & $390$
&
\multicolumn{3}{c|}{}
\\
$p133$
& $1.4$ 
& $1.4\cdot10^{-4}$
& $p333$ 
& $1.3\cdot10^3$ & $1.6\cdot10^5$
&
\multicolumn{3}{c|}{}
\\
$p211$
& \multicolumn{2}{c|}{$4.9(7.3)\cdot10^{-10}$}
&\multicolumn{3}{c|}{}
&\multicolumn{3}{c|}{}
\\\cline{1-9}
%%%%%%%%%%%%%%%%%%%%%%%%%%%%%%%%%%%%%%%%%%%%%
\multicolumn{9}{|c|}{\large$\calO_{\bar{e}ddDd}$}
\\\cline{1-9}
$1111$ &\multicolumn{2}{c|}{$9.2\cdot10^{4}$}
&$2111$ &\multicolumn{2}{c|}{$450$}
&$3111$ &\multicolumn{2}{c|}{$27$}
\\
$1112$ &\multicolumn{2}{c|}{$2.0\cdot10^{4}$}
&$2112$ &\multicolumn{2}{c|}{$95$}
&$3112$ &\multicolumn{2}{c|}{$5.7$}
\\
$1113$ &\multicolumn{2}{c|}{$2.4\cdot10^{4}$}
&$2113$ &\multicolumn{2}{c|}{$110$}
&$3113$ &\multicolumn{2}{c|}{$6.8$}
\\
$1122$ &\multicolumn{2}{c|}{$4.1\cdot10^{4}$}
&$2122$ &\multicolumn{2}{c|}{$200$}
&$3122$ &\multicolumn{2}{c|}{$12$}
\\
$1123$ &\multicolumn{2}{c|}{$4.8\cdot10^{4}$}
&$2123$ &\multicolumn{2}{c|}{$230$}
&$3123$ &\multicolumn{2}{c|}{$14$}
\\
$1133$ &\multicolumn{2}{c|}{$3.5\cdot10^{6}$}
&$2133$ &\multicolumn{2}{c|}{$1.7\cdot10^4$}
&$3133$ &\multicolumn{2}{c|}{$1.0\cdot10^3$}
\\
$1222$ &\multicolumn{2}{c|}{$1.4\cdot10^{10}$}
&$2222$ &\multicolumn{2}{c|}{$6.9\cdot10^7$}
&$3222$ &\multicolumn{2}{c|}{$4.1\cdot10^6$}
\\
$1223$ &\multicolumn{2}{c|}{$1.2\cdot10^{7}$}
&$2223$ &\multicolumn{2}{c|}{$5.6\cdot10^4$}
&$3223$ &\multicolumn{2}{c|}{$3.3\cdot10^3$}
\\
$1233$ &\multicolumn{2}{c|}{$1.4\cdot10^{7}$}
&$2233$ &\multicolumn{2}{c|}{$6.8\cdot10^4$}
&$3233$ &\multicolumn{2}{c|}{$4.0\cdot10^3$}
\\
$1333$ &$2.3\cdot10^{13}$
&$6.9\cdot10^{14}$
&$2333$ &$1.1\cdot10^{11}$
&$3.4\cdot10^{12}$
&$3333$ &$6.7\cdot10^{9}$
&$2.0\cdot10^{11}$
\\\cline{1-9}
\end{tabular} }
\caption{Upper limits on the dimensionless WCs for all possible generation combinations of the six dim-7 SMEFT BNV operators, derived from two-body nucleon decay processes and incorporating complete one-loop RG running effects. The numbers in parentheses indicate the results without including RG running effects.}
\label{tab:limit_dim7_Ops}
\end{table}  

%%%%%%%%%%%%%%%%%%%%%%%%
\section{An illustrative example}
\label{app:example}
%%%%%%%%%%%%%%%%%%%%%%%%

To illustrate the procedure and clarify the resulting constraints, we work in the up-quark flavor basis and consider the operator $\calO_{\bar{L}dQQ\tilde{H}}^{3121}$
as an example, assuming that only its WC $C_{\bar{L}dQQ\tilde{H}}^{3121}$ is nonzero  at $\Lambda_{\tt NP}=10^8$ GeV.
From \cref{tab:matching_EW}, we see that, in the absence of RG running effects, 
it cannot induce leading-order LEFT operators contributing to nucleon decay processes listed in \cref{tab:exp_bound} due to the appearance of the heavier $\tau$ lepton.
However, once the SMEFT RG running is taken into account, this operator can induce additional LEFT operators at $\Lambda_{\tt EW}$ that contribute to nucleon decay modes and therefore is subject to stringent constraints.

To see the effect, we assume that only $C_{\bar{L}dQQ\tilde{H}}^{3121}$ is nonvanishing at $\Lambda_{\tt NP}=10^8$ GeV, and then solve the complete SMEFT RG equations using $\tt{D7RGESolver}$~\cite{Liao:2025lxg}.
In the unit of $C_{\bar{L}dQQ\tilde{H}}^{3121}(\Lambda_{\tt NP})$, the resulting SMEFT WCs at $\Lambda_{\tt EW}$ are
{\small
\begin{verbatim}
**EFT:** `SMEFT`
| WC name | Value |
|--------------|--------------------------------------------------|
| `LdQQH_3121` | (1.3553462840737287-6.796539802696721e-21j) |
| `LdQQH_3112` | (0.13738480101984119-1.455115193168033e-20j) |
| `LdQQH_3113` | (7.386078206860526e-07+4.937554614674685e-12j) |
| `eQddH_3113` | (-6.134846454443941e-07+9.120319818869216e-17j) |
| `eQddH_3112` | (-2.7359067568382057e-07+9.199966482392824e-12j) |
| `LdQQH_3131` | (-2.6368470976782523e-07-1.7744582663788962e-12j) |
| `LdQQH_3123` | (-5.012628302531261e-08+1.6247510605911527e-07j) |
| `LdudH_3122` | (3.7416242123788585e-08+6.396357782421948e-16j) |
| `LdudH_3123` | (9.234958483364565e-09-2.983702324746525e-08j) |
| `eQddH_3212` | (2.8894570711364855e-08-1.5135393431169674e-15j) |
| `eQddH_3213` | (7.1316652456395e-09-2.3041531491577803e-08j) |
| ... | ... |
\end{verbatim}}
\normalsize
\noindent
From the output results, 
it can be seen that, except for $C_{\bar{L}dQQ\tilde{H}}^{3121}$ itself,  
the next largest nonzero WC at $\Lambda_{\tt EW}$ is
\begin{align}
    C_{\bar{L}dQQ\tilde{H}}^{3112}(\Lambda_{\tt EW})&\approx 0.137  C_{\bar{L}dQQ\tilde{H}}^{3121}(\Lambda_{\tt NP})\;. 
\end{align}
This is generated through the following SMEFT RGE~\cite{Liao:2016hru}
\begin{align}
    \frac{d}{d \ln\mu} C_{\bar{L}dQQ\tilde{H}}^{3112}=-\frac{3g_2^2}{16\pi^2}C_{\bar{L}dQQ\tilde{H}}^{3121}+\cdots\;.
\end{align}
The nonzero SMEFT WC $C_{\bar{L}dQQ\tilde{H}}^{3112}$ at $\Lambda_{\tt EW}$ can then be matched to the following LEFT WC contributing to nucleon decays (see \cref{tab:matching_EW}),
\begin{align}
    C_{\bar{\nu}dsu}^{\tR\tL,\tau}(\Lambda_{\tt EW})&=\frac{v}{\sqrt{2}}V_{22}C_{\bar{L}dQQ\tilde{H}}^{3112}(\Lambda_{\tt EW})
    \approx 23.2\times[C_{\bar{L}dQQ\tilde{H}}^{3121}(\Lambda_{\tt NP}){\rm{GeV}}^3]\times{\rm{GeV}}^{-2}\;.
\end{align}
Substituting $C_{\bar{\nu}dsu}^{\tR\tL,\tau}(\Lambda_{\tt EW})$ into  \cref{tab:B2lM_vertex} and \cref{eq:GammaN2lM}, we find that the two nucleon decay modes involving a neutrino and a kaon are induced, and their corresponding inverse decay widths are 
\begin{align}
\Gamma^{-1}(p\to \nu_\tau K^+)\approx 6.61\times 10^{33}\left|\frac{1.9\times 10^{-32}\,\rm{GeV}^{-3}}{C_{\bar{L}dQQ\tilde{H}}^{3121}(\Lambda_{\tt NP})}\right|^2\ {\rm{yr}},
\quad
\Gamma^{-1}(n\to \nu_\tau K^0)\approx 7.8\times 10^{32}\left|\frac{2.2\times 10^{-32}\,\rm{GeV}^{-3}}{C_{\bar{L}dQQ\tilde{H}}^{3121}(\Lambda_{\tt NP})}\right|^2\ {\rm{yr}}.
\end{align}
Comparing them with the experimental bounds listed in \cref{tab:exp_bound}, we find that the decay mode $p\to \nu_\tau K^+$ imposes the strongest constraint, with 
$C_{\bar{L}dQQ\tilde{H}}^{3121}(\Lambda_{\tt NP})\lesssim 1.9\times10^{-32}\,\rm{GeV}^{-3}$
as we show in \cref{tab:limit_dim7_Ops}. 

\twocolumngrid
\bibliography{references_paper}{}
\bibliographystyle{utphys}

%%%%%%%%%%%%%%%%%%%%%%%%
\end{document}